\documentclass[12pt,preprint]{aastex}
\usepackage[numberedappendix]{emulateapj5}
\shorttitle{NESTED AND SINGLE BARS}
\shortauthors{LAINE ET AL.}
\begin{document}

\submitted{TO APPEAR IN THE ASTROPHYSICAL JOURNAL.}
\title{NESTED AND SINGLE BARS IN SEYFERT AND NON-SEYFERT GALAXIES}

\author{Seppo Laine\altaffilmark{1}}
\affil{Department of Physics and Astronomy, University of Kentucky,
    Lexington, KY 40506-0055}
\email{laine@stsci.edu}

\author{Isaac Shlosman\altaffilmark{2,3}}
\affil{Joint Institute for Laboratory Astrophysics, University of Colorado,
    Box 440, Boulder, CO 80309-440}
\email{shlosman@pa.uky.edu}

\author{Johan H. Knapen}
\affil{Isaac Newton Group of Telescopes, Apartado 321, E-38700 Santa Cruz de 
la Palma, Spain\\ and Department of Physical Sciences, 
University of Hertfordshire, Hatfield, Herts. \\AL10 9AB, United Kingdom}
\email{knapen@ing.iac.es}

\and

\author{Reynier F. Peletier}
\affil{School of Physics and Astronomy, University of Nottingham, Nottingham,
NG7 2RD, United Kingdom}
\email{Reynier.Peletier@nottingham.ac.uk}

\altaffiltext{1}{present address: Space Telescope Science Institute, 3700 San
Martin Drive, Baltimore, MD 21218}
\altaffiltext{2}{JILA Visiting Fellow}
\altaffiltext{3}{permanent address: Department of Physics and Astronomy, 
University of Kentucky, Lexington, KY 40506-0055}
    
\begin{abstract} We analyze the observed properties of nested and single
stellar bar systems in disk galaxies. The 112 galaxies in our sample comprise
the largest matched Seyfert vs. non-Seyfert galaxy sample of nearby galaxies
with complete near-infrared or optical imaging sensitive to lengthscales
ranging from tens of pc to tens of kpc. The presence of bars is deduced by
fitting ellipses to isophotes in {\it HST} $H$-band images up to 10$''$ radius,
and in ground-based near-infrared and optical images outside the $H$-band
images. This is a conservative approach that is likely to result in an
underestimate of the true bar fraction. We find that a significant fraction of
the sample galaxies, 17\% $\pm$ 4\%, has more than one bar, and that 28\% $\pm$
5\% of barred galaxies have nested bars. The bar fractions appear to be stable 
according to reasonable changes in our adopted bar criteria. For the nested 
bars, we detect a clear division in length between the large-scale (primary)
bars and small-scale (secondary) bars, both in absolute and normalized (to the
size of the galaxy) length. We argue that this bimodal distribution can be
understood within the framework of disk resonances, specifically the inner
Lindblad resonances (ILRs), which are located where the gravitational potential
of the innermost galaxy switches effectively from 3D to 2D. This conclusion is
further strengthened by the observed distribution of the sizes of nuclear rings
which are dynamically associated with the ILRs. While primary bars are found to
correlate with the host galaxy sizes, no such correlation is observed for the
secondary bars. Moreover, we find that secondary bars differ morphologically
from single bars. Our matched Seyfert and non-Seyfert samples show a
statistically significant excess of bars among the Seyfert galaxies at
practically all lengthscales. We confirm our previous results that bars are
more abundant in Seyfert hosts than in non-Seyferts, and that Seyfert galaxies
always show a preponderance of ``thick'' bars compared to the bars in
non-Seyfert galaxies. Finally, no correlation is observed between the presence
of a bar and that of companion galaxies, even relatively bright ones. Overall,
since star formation and dust extinction can be significant even in the
$H$-band, the stellar dynamics of the central kiloparsec cannot always be 
revealed reliably by the use of near-infrared surface photometry alone. 
\end{abstract}


\section{INTRODUCTION}

While a substantial effort has been spent on understanding the prevalence and
properties of kpc-scale bars, little is known about bars on sub-kpc scales.
These ``inner'' or ``nuclear'' bars were first discovered as optical isophote
twists in the central regions of barred galaxies, e.g., de Vaucouleurs (1974), 
Sandage \& Brucato (1979) and Kormendy (1982), and interpreted as triaxial
bulges  of barred galaxies. Later ground-based studies at higher resolution
have revealed more galaxies with nuclear bars, lying inside large galactic bars
(e.g. Buta \& Crocker 1991, 1993; Shaw et  al. 1993, 1995; Knapen et al. 1995a;
Wozniak et al. 1995; M\"{o}llenhoff et al. 1995; Friedli et al. 1996; 
Jungwiert, Combes, \& Axon 1997; Mulchaey \& Regan 1997; Elmegreen et al. 1998;
Jogee, Kenney, \& Smith 1998, 1999; Knapen, Shlosman, \& Peletier 2000, hereafter
KSP; M\'arquez et al. 2000; Greusard et al. 2000). Some of the earlier studies
have been summarized by Buta \& Combes (1996).

Although the first detections of sub-kpc bars were made in stellar light, these
objects can contain arbitrary fractions of gas, and in extreme cases  can be
dynamically dominated by molecular gas, as evident in their detection
in interferometric 2.6~mm CO emission and in the near-infrared
(NIR) lines of H$_2$ emission (e.g. Ishizuki et al. 1990; Devereux, Kenney, \&
Young 1992; Forbes et al. 1994; Mirabel et al. 1999; Kotilainen et al. 2000;
Maiolino et al. 2000). CO observations have a rather low spatial resolution
(at best just below 1$''$ FWHM), but do allow the determination of the offset
angle between the large-scale stellar and the small-scale gaseous bar. It is 
not yet clear whether stellar-dominated and gas-dominated sub-kpc bars have a
common origin or describe a concurrent phenomenon. In this paper we focus on
stellar bars, analyzing their properties based on NIR and optical starlight.

The high spatial resolution capability provided by the {\it Hubble Space
Telescope (HST)} has enabled studies of galaxy centers with about 0\farcs 1
resolution. More embedded nuclear bars have been detected in these observations
(e.g., Erwin \& Sparke 1999a,1999b; Regan \& Mulchaey 1999; Martini \& Pogge
1999; Colina \& Wada 2000; Emsellem \& Ferruit 2000; van den Bosch \& Emsellem
2000) using various techniques, but most of these papers only discuss one
galaxy, apart from  those by Regan \& Mulchaey (1999) and  Martini \& Pogge
(1999), where 12 and 24 galaxies were considered, respectively. 

Because the existence of a nested bar system is  intrinsically a time-dependent
phenomenon, no single-periodic orbits, which form the ``backbone'' of a barred
galaxy, can dominate the stellar dynamics. Maciejewski \& Sparke (2000) discuss
a special class of double bar orbits which close after a number of rotations,
but the exact fraction of the total phase space occupied by these orbits is
unknown, and whether they are significantly populated is unclear. Such orbits
can only host stars: because all these orbits intersect, they are clearly not
suitable for the gas and  cannot support the steady-state gas motions. Because
of these, and additional reasons, offset dust lanes, which characterize the gas
motions in the large stellar bars, do not form in nuclear bars, and cannot be
used as indicators of bar presence (Shlosman \& Heller 2001). The sparseness of
nuclear bars in the recent NIR snapshot survey by Regan \& Mulchaey (1999) and
Martini \& Pogge (1999), who attempted to find sub-kpc bars based on the offset
dust lanes, should therefore not be surprising, as these studies postulated
identical gas flows in large- and small-scale bars.

This work is the first attempt to determine the statistics of nested bars in
disk galaxies and to compare nested bars with single bars in Seyfert and
non-Seyfert host galaxies. For this we have taken a large sample of nearby
spiral galaxies with available {\it HST} archive NICMOS images in the $H$-band,
added ground-based data of the outer disks, and analyzed all the images for
the  presence of bars using fits of ellipses to isophotes. This is the largest 
high spatial resolution sample analyzed so far. As a by-product, we are able to
test our previous results (KSP) on the large-scale bar fractions  in Seyfert
and non-Seyfert galaxies, with improved  statistical significance.

We give the definitions of the bar concepts that we use throughout our paper in
Section 2 and describe our samples and the analysis of the images in Section 3.
Results on nested and single bars are presented in Section 4 and the
differences between Seyfert and non-Seyfert bar properties are given in Section
5. Discussion and conclusions are provided in Sections 6 and 7, respectively.

\section{DEFINING ASPECTS OF NUCLEAR BARS}

We define nested bar systems as those with more than one bar, stellar or
gaseous, and focus here on purely stellar bars. Although the terms ``nuclear
bar'' or ``inner bar'' have been used  in the literature to refer to bars other
than the largest bar in a system, these terms are loosely defined. Overall they
refer to bars lying in scarcely resolved nuclear regions of disk galaxies,
generally within $\sim 1$~kpc. Naturally the names do not reflect any special
physical properties of these bars. To avoid further ambiguities with bar
definitions, we use the following notation.

We first distinguish between large- and small-scale bars in double-barred
systems. The former are referred to as ``primary'', while the latter are called
``secondary''. The theoretical rationale behind these definitions is that
secondary bars are believed to form as a result of radial gas inflow due to the
large-scale bar and, therefore, are expected to be confined within the inner
Lindblad resonance(s) (hereafter ILRs), which naturally limit(s) their size to
$\sim 1$~kpc (Shlosman, Frank, \& Begelman 1989). Within this framework,
molecular gas accumulation in the vicinity of the ILRs is susceptible to a
global gravitational (bar) instability which will affect stars as well, by
dragging them along. Secondary bars would then form with a pattern speed which
is higher than that of the primary bar. This view was supported by Pfenniger \&
Norman (1990) who, using weakly dissipative equations of motion for a test
particle, analyzed the properties of double-barred galaxies. In particular,
they argued that the corotation radius of the secondary bar should coincide
with the ILR of the primary bar in order to reduce the fraction of chaotic
orbits in the resonance neighborhood. This suggestion was made on purely
theoretical grounds. It means that the ILRs serve as a dynamical separator
between primary and secondary bars, a point we address observationally in the
current paper. Finally, bars in galaxies which host only one bar, are called
``single bars'', independent of their physical length. 

\section{SAMPLE AND DATA ANALYSIS}

\subsection{Sample}

Our Seyfert sample consists of most of the Seyfert galaxies in the local
universe ($v_{\rm hel}$ $<$ 6000 km~s$^{-1}$) that have been observed in the F160W
($H$) band with {\it HST}. Most of the Seyfert galaxies come from the 
relatively large samples of Mulchaey (part of which was published by Regan \&
Mulchaey 1999), Stiavelli (Seigar et al.  2000), Pogge (Martini \& Pogge 1999)
and Peletier (Peletier et al.  1999). A few well-known Seyferts were added to
our Seyfert sample: NGC 1068 (Thompson \& Corbin 1999), NGC 3227 (Quillen et
al. 1999),  NGC 4151  (R. Thompson, unpublished), NGC 5548 (M. Rieke,
unpublished), and NGC 7469 (Scoville et al. 2000). As in KSP, we removed highly
inclined galaxies by requiring that the  apparent axial ratio, as obtained from
the RC3 (de Vaucouleurs et al. 1991),  had to be greater than 0.45, since the
detection of nonaxisymmetric structures in highly inclined galaxies is
problematic. We included galaxies which have been classified in RC3  with
Hubble types S0 to Sc because there are very few Seyferts of later types. Since
we are interested in the central kpc of disk galaxies, we did not discard
interacting galaxies, except when the interaction was accompanied by a strong
morphological distortion. Note that the effect of interactions on the galactic
morphology in the central kpc generally is not significant (unless it is
a full-fledged merger) because of short relaxation timescales in the
circumnuclear region. We further discuss the  influence of companion galaxies
in Section 6. Galaxies which had been so badly centered that their nucleus was
lying close to the edge of the NICMOS field were excluded. This left us with a
final sample of 56 Seyfert galaxies. We then matched the Seyfert sample by a
control sample of 56 non-Seyfert galaxies, also observed in the F160W band of 
NICMOS. 

\subsection{Sample Matching}

We constructed the control sample to have a similar distribution to the Seyfert
sample in the following four parameters: absolute $B$ magnitude, distance, axis
ratio, and morphological type. Our control galaxies were selected from a sample
of 95 non-Seyfert galaxies for which {\it HST} $H$-band images existed.  To
match the distributions of the non-Seyfert galaxies to those of the Seyfert
galaxies in the afore-mentioned four parameters, we divided the Seyferts into
two magnitude bins, two distance bins, five axis ratio bins, and six
morphological type bins. We then eliminated non-Seyfert galaxies in bins that
had an excess of them compared to the Seyferts. This procedure was repeated
until a match was found that produced the smallest total difference in the bin
distributions between the Seyfert and non-Seyfert galaxies. In the absolute $B$
magnitude, which had been  derived from the apparent $B_{T,0}$ magnitude in RC3
and the distance, we split the samples at magnitude $-20.4$, which divides the
samples into two bins with roughly equal numbers of galaxies. Changes in this
dividing magnitude of more than 0.1 would produce a very different (and
unequal) division of  the galaxies in the two magnitude bins. The distance was
obtained  from Tully's Nearby Galaxies Catalog (1988) or from the heliocentric
velocity  using the Hubble law and a Hubble constant of
75~km~s$^{-1}$~Mpc$^{-1}$. We divided the samples into two distance bins with
the  bin separation at 28 Mpc. Changing the dividing boundary by more than 2
Mpc would produce very different (and uneven) partitions in the two bins. For
the axis ratio, the values separating the bins are 0.6, 0.7, 0.8, and 0.9. For
the morphological type, we used bins in the numerical Hubble Type~T given in
RC3. Graphical representations and tables illustrating our sample matching are
given in Appendix A.

\subsection{Data Reduction}

We chose to analyze $H$-band images only. $K$-band images would likely be more
reliable since they suffer less from the disturbing effects of dust
extinction,  but unfortunately $K$-band {\it HST} images were not available for
the majority of our sample galaxies. We started with the data produced by the
NICMOS pipeline, as available in the {\it HST} archive. Further reduction involved
the masking of artifacts from the NICMOS images, and sometimes the removal of
an additional pedestal level, to make the background flat.  The artifacts often
included the coronagraphic hole of the NICMOS camera and involved masking of
the central columns of the images which were mosaiced by the NICMOS calibration
pipeline. Nearby stars were also masked out. In case of bright point source
nuclei, we generated a PSF with the Tiny Tim software and ran a few iterations
of Lucy deconvolution. In extreme cases of diffraction, we used the IRAF task
CPLUCY to perform the deconvolution in the central region. This deconvolution
eliminated the original diffraction rings completely. The outer spikes of the
PSF were manually masked out.

\subsection{Bar Detection and Classification}

Since we aim to detect all the bars in the sample, whether they are
secondary, primary or single bars, the {\it HST} data have been extended to
larger radii. Ideally, we preferred to have NIR images with a large
field of view, thus covering the galaxies completely. Since such images,
unfortunately, were not available, we applied a different solution. Just
beyond the range of the NICMOS images, we use data from the 2MASS all-sky
survey. Although it is not a very deep survey and has a relatively low spatial
resolution (about 2$''$), its $H$ band is compatible, and it allows us to
cover the radial range between typically 5$''$ and $30''$. Further out we use
optical images, either CCD-images from various data archives or digitized sky
survey (DSS) images. Although optical images are affected considerably more by
dust extinction and star formation, thus making bar classification much more
difficult, most of the dust is probably found in the inner regions (e.g.,
Peletier et al. 1995; Giovanelli et al. 1994), and not many bars will be
missed.

We used the GALPHOT package (see J{\o}rgensen, Franx, \& Kjaergaard 1992), run
under the IRAF\footnote{IRAF is distributed by the National Optical Astronomy
Observatories, which are operated by the Association of Universities for
Research in Astronomy, Inc., under cooperative agreement with the National
Science Foundation.} environment, to fit elliptical isophotes to the galaxy
images. We also followed closely the procedure given in Peletier, Knapen, \&
Shlosman (1999). The center of the isophotes was given an initial guess based
on the location of the peak in the image, but the fitting program had the
freedom to move the center of the fitted ellipses. We fitted ellipses usually
in multiplicative (1.1) radial increments. The fitted position angle in the
images was transformed to true position angle (measured east of north on the
sky) by using the header information in the {\it HST} images. A few of our
sample galaxies were not well centered in the field of view of the camera and,
therefore, it was not possible to reliably fit these galaxies to the edge of
the field of view, because only a small fraction of the ellipses would lie on
the image (the NICMOS Camera 2, which was used for most of the images,  has a
field of view of 19 \farcs 2 $\times$ 19 \farcs 2).  Similarly, ellipses were
fitted to the images from other sources (2MASS, DSS, etc.). The fitted ellipses
were then deprojected using a two-dimensional deprojection. For this we
assumed that the outer parts of the galaxies are flat circular disks,  with
the inclination given by the axis ratio in NED.

Next, we describe our method of bar detection on all possible lengthscales in
the images. To classify a galaxy as barred, we use the criteria set in KSP,
namely, a bar is revealed by a significant rise in ellipticity ($1-b/a$),
followed by a significant fall, while the position angle of the major axis of
the fitted ellipse is roughly constant. To quantify this a little further, we
require that the ellipticity variation has an amplitude of at least 0.1
(increase and decrease) while the position angle varies by less than 20$\degr$.
To be conservative we  did not follow the second criterion of KSP, namely,
position angle twists of more than 75$\degr$ accompanied by ellipticities above
the 0.1 level. We have checked that the bar fractions are stable against
reasonable changes in the bar criteria, such as a change of 20\% in ellipticity
amplitude requirement and a change of 50\% in the constancy of the
position angle  requirement. The bar length was defined to be the radius where
the fitted bar ellipticity peaks.  The bar ellipticity was defined as the
maximum ellipticity of a detected bar. Tabulated properties of the bars and
graphical representations of the deprojected ellipse fits are given in Appendix
B.

\section{OBSERVED NESTED AND SINGLE BAR PROPERTIES}

\subsection{Nested Bars}

\subsubsection{Overall Statistics}

In total we find that 69 of our 112 galaxies have at least one bar (62\% $\pm$
5\%). We use Poisson statistics to give an uncertainty estimate of our numbers,
estimated from the formula $\sigma$ =  $\sqrt{f(1-f/N)}$, where $f$ is the
quantity that is measured and $N$ is the sample size in which this quantity is
searched ($f=69$ and $N=112$ above). There are 12 (21\% $\pm$ 5\%) Seyfert and
7 (13\% $\pm$ 4\%) non-Seyfert galaxies in our samples which have nested bar
systems.  These include two triple-barred systems among the Seyferts. In
triple-barred systems we classify the outermost bar as a primary bar and the
two innermost bars as secondaries.  Altogether we have found secondary or
primary bars in 19 (17\% $\pm$ 4\%) galaxies. This represents the bar fraction
in spiral galaxies which have a morphological type distribution of the Seyfert
hosts in our sample, making our results somewhat biased toward early-type
spirals. 

\subsubsection{Bar Size Distribution in Nested Systems}

Figure~1 shows the distribution of nested bar sizes, normalized by galactic
diameter ($D_{25}$). The data were divided into two groups which were split at
a value which minimizes the overlap between primary and secondary bars in
nested bar systems. We find that the minimal overlap between the distribution
of normalized primary and secondary bar sizes for both the Seyfert and
non-Seyfert galaxies occurs at a (normalized) bar length of $l_{\rm crit} \approx
0.06$. As discussed in Section 6.1, $l_{\rm crit}$ also exists in the physical bar
length domain and corresponds to  $\approx$ 1.6 kpc. Interestingly, without the
triple-barred systems only one primary bar lies on the ``wrong'' side of the
dividing line in the normalized diagram. This is the first time that such a
clear separation of primary and secondary bar lengths has been shown
observationally. Because our samples include all the Hubble types from S0 to
Sc, the minimal overlap between the two bar classes means that our result
stands regardless of the morphological class of the galaxy. 

\includegraphics[width=3.0in]{fig1cor.eps}
\figcaption{Distribution of normalized (a.) and physical (b.) primary 
(cross-hatched) and secondary (blank) bar sizes. The top panels show Seyfert 
galaxies, the middle panels non-Seyfert galaxies, and the bottom panels display
the totals. The bar lengths were normalized by the host galaxy diameter 
$D_{25}$ and the resulting values were divided into two groups,  l$<$ $l_{\rm
crit}=0.06$ and l$>$ $l_{\rm crit}=0.06$. In physical units the critical 
dividing length is 1.6 kpc. The critical dividing bar lengths $l_{\rm crit}$
were chosen to minimize the overlap between the two groups.  \label{fig1}}
\vskip0.1in

\subsubsection{Ellipticities of Nested Bars}

Figure~2 shows the ellipticity distributions of nested bars in our Seyfert and
non-Seyfert subsamples. To minimize the uncertainties we have divided all bars
into two groups based on their ellipticities, $\epsilon \leq 0.45$ and 
$\epsilon > 0.45$. The results are not overly sensitive to the exact position
of this boundary. It is clear from this figure that secondary bars have a
larger fraction of lower ellipticities than primary bars, both among Seyferts
and among non-Seyferts. In fact, among the Seyfert double-barred systems, 80\%
$\pm$ 13\% have a higher outer bar ellipticity than inner bar ellipticity, and
71\% $\pm$ 17\% of non-Seyfert double-barred systems have a higher outer bar
ellipticity. The simplest explanation for the prevalence of less elliptical
secondary bars is that their ellipticity is diluted by the light from the
galactic bulge, which is expected to be rounder than the underlying bar.  

Figures 3a,b and 4a,b display the physical and normalized (divided by the $D_{25}$
diameter) bar lengths vs. the deprojected ellipticities of the bars.  There is
a sharp increase in ellipticity towards large-scale bars, especially  for
Seyfert galaxies. In fact, this increase is evident for all sizes in excess of 
$l_{\rm crit}$ in normalized or physical bar lengths.  A similar correlation,
namely that strong bars are long, has been seen earlier in later Hubble type
galaxies (Martinet \& Friedli 1997; M\'arquez et al. 2000). Here we extend this
correlation between the bar ellipticity and its length to earlier Hubble types.

\vskip0.1in

\includegraphics[width=2.5in]{fig2cor.eps}
\figcaption{Distribution of bar ellipticities in nested and single-barred
Seyfert and non-Seyfert host galaxies. Each pair of columns represent bars 
with $\epsilon \leq 0.45$ (black; ``fatter'' or ``weaker'' bars) and bars 
with $\epsilon > 0.45$ (hatched; ``leaner'' or ``stronger'' bars). 
\label{fig2}} 
\vskip0.1in

\subsection{Comparison of Single and Nested Bars}

Can the single bars be meaningfully divided into two groups based on their 
lengths, in analogy with nested bars, where there is a fairly sharp division
between secondary and primary bars in terms of their length? In other words, do
the  small-scale {\it single} bars exhibit properties similar to the secondary
bars, which might hint about a common origin.  To test this conjecture we
compared the properties of nested and single bars. Figures~3 and 4 show that
if the division at $l_{\rm crit}$ is used, only a relatively small fraction of 
single bars have lengths less than $l_{\rm crit}$ (7/29 among the Seyferts and
8/21 among the non-Seyferts). 

The ratio of the number of secondary to nested (primary plus secondary) bars 
is 54\% among the Seyferts, (including the triple-barred systems) and
(obviously) 50\%  among the non-Seyferts. While the single bars predominantly
lie outside $l_{\rm crit}$ in both of our subsamples (Figs. 3c,d, and 4c,d),
the fraction of small-scale  ($l$ $< l_{\rm crit}$) single bars among all the
single bars is only 24\% $\pm$ 8\% in the Seyfert sample and 38\% $\pm$ 11\% in
the non-Seyfert sample. Thus the fraction of small-scale single bars is
significantly smaller than the corresponding fraction of secondary bars. This
is the first indication that small single bars and secondary nested bars
may have a different origin.

The second indication of this dissimilarity is that the ellipticity of single
bars appears to be distributed differently from that of nested bars (Fig.~2).
Single bars have a higher fraction of large ellipticities not only when
compared to nested bars as a whole, but also when comparing single bars
to primary bars alone, as can be seen in Fig.~2. While the fraction of nested
bars with an ellipticity $< 0.45$ is greater than or equal to the fraction of
nested bars with an ellipticity $>$ 0.45 among the Seyferts and non-Seyferts, 
the majority of single bars have ellipticities $> 0.45$. This may be partly
related to the fact that the majority of single bars are large-scale bars and,
therefore, have larger ellipticities (Section 4.1.3).

Finally, we inspect the Hubble type distribution of the host galaxies of the
various bar classes. Figure 5 compares the fractions of early-type (S0--Sa)
galaxies with bars to the fraction of late-type (Sb--Sc) galaxies with bars,
separately for Seyfert and non-Seyfert samples, and nested and single bars.
Nested bars prefer later Hubble types whereas the single
bars occupy about equal fractions among early- and late-type galaxies. Although the  numbers are
small, this is possibly yet another indication of the different origin of
nested and single bars. The most striking result in this figure is that nested
bar systems in non-Seyfert  galaxies only occur in late Hubble types (Sb--Sc).
This cannot be due to a lack of early-type galaxies in our non-Seyfert sample
since we matched the two samples in morphological type, but of course it can be
a result of small number statistics.

\begin{figure*}
\centering
\includegraphics[width=6in]{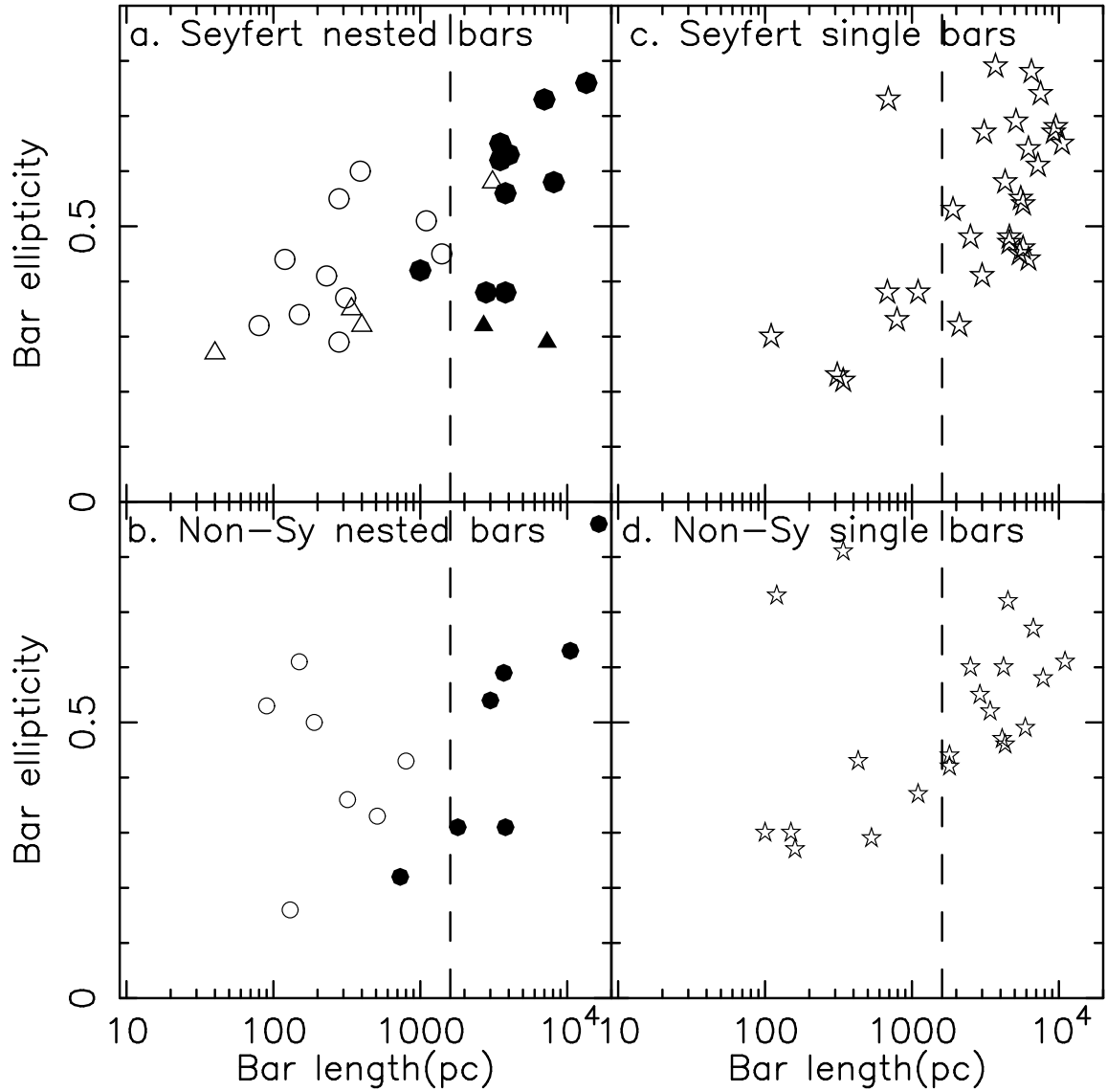}
\caption{Deprojected physical bar lengths for all the bars in our samples, 
separately for Seyfert and non-Seyfert galaxies. The y-axis is the ellipticity 
after a two-dimensional deprojection. a. Primary and secondary bars in the
Seyfert sample. Primary bars are shown with filled circles, secondary bars with
open circles. Triangles denote bars in triple-barred galaxies. b. Same as
a. but for non-Seyfert galaxies. c. Single bars in the Seyfert sample. d.
Single bars in the non-Seyfert sample.\label{fig3}}
\end{figure*}

\begin{figure*}
\includegraphics[width=6in]{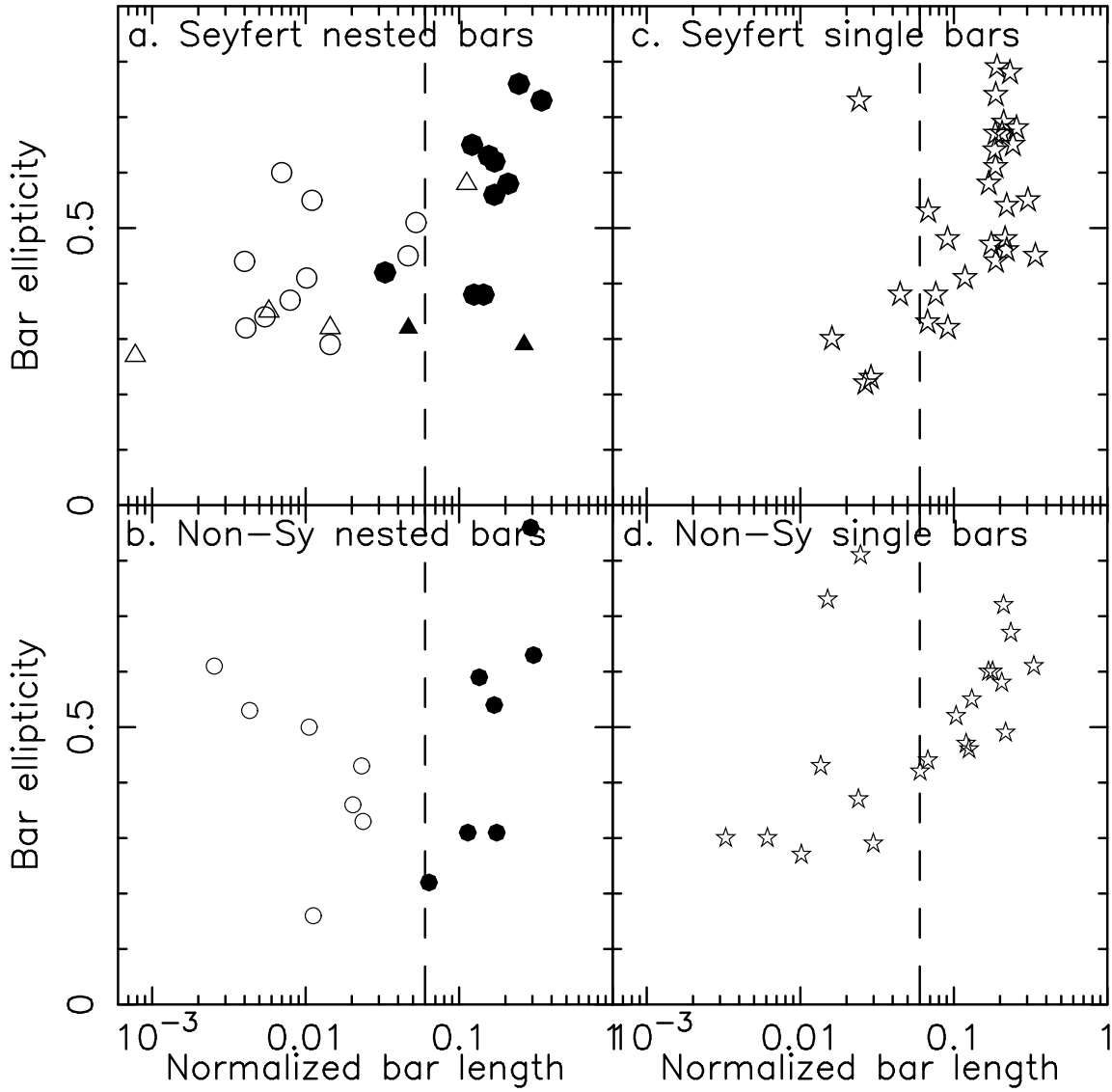}
\caption{As Fig 3, but now showing normalized deprojected bar lengths listed 
in Tables 6 and 7. 
\label{fig4}}
\end{figure*}

\section{COMPARISON OF BARS AMONG SEYFERT AND NON-SEYFERT GALAXIES}

\subsection{Bar Fraction in Seyfert and Non-Seyfert Host Galaxies}

Morphological differences between Seyferts and non-Seyferts, in general, and
different bar fractions, in particular, have been sought in order to understand
the fueling mechanism(s) of central stellar and non-stellar activity in AGNs
(e.g. Adams 1977; Simkin, Sue, \& Schwartz 1980; Balick \& Heckman 1982;
MacKenty 1989; and more recent studies by Moles, M\'arquez, \& P\'erez 1995; Ho,
Filippenko, \& Sargent 1997a). The early surveys, but also the more recent ones,
suffer from an absence of a properly matched control sample, from low
resolution, from being conducted in the optical band where stellar bars are
more difficult to detect, or from adopting the RC3 classification. When such
studies were performed in the NIR, the fraction of barred galaxies increased by
a factor of two at least compared to the fraction with SB notation in RC3, but
the difference between Seyferts and non-Seyferts has been found to be
statistically  insignificant (Mulchaey \& Regan 1997) or marginally significant
(KSP). The newest careful NIR observations have revealed an even larger
fraction of bars (Grosb{\o}l 2001).

\includegraphics[width=3.0in]{fig5cor.eps}
\figcaption{The fractions of Seyfert (top), non-Seyfert (middle) and 
total (bottom) galaxies with nested or single bars, divided into early
(S0--Sa; black column) and late (Sb--Sc; cross-hatched column) Hubble classes. 
The morphological classifications were taken from NED.\label{fig5}}
\vskip0.1in

To detect bars, KSP used NIR imaging data for the CfA sample (Huchra \& Burg
1992), observed with sub-arcsecond resolution, and applied a set of
well-defined and objective criteria. A difference between the barred fraction
in Seyferts and non-Seyferts was found, but only  at a 2$\sigma$ significance
level (Table 1). Because the current sample is larger than that of KSP, we can
perform the test again, improving its statistical significance. Since we
include the {\it HST} archive data, we are also able to detect smaller bars
than KSP. 

\begin{table*}
\caption{Comparison of the bar statistics from KSP to the current
paper.\label{tab1}}
\begin{tabular}{lccccc}
\tableline
\tableline
Sample & Size (Sy, control) & \multicolumn{2}{c}{Bar fraction: Seyferts} &
\multicolumn{2}{c}{Bar fraction: control}\\
\tableline
&& $N$ & Percentage & $N$ & Percentage \\
KSP & 29, 29 & 23/29 & 79\% $\pm$ 8\% & 17/29 & 59\% $\pm$ 9\% \\
This paper & 56, 56 & 41/56 & 73\% $\pm$ 6\% & 28/56 & 50\% $\pm$ 7\%\\
\tableline
\end{tabular}
\end{table*}

Of the 56 galaxies in our current Seyfert sample, we find that 41 are barred
(73\% $\pm$ 6\%), whereas of our 56 control galaxies, only  28 (50\% $\pm$ 7\%)
are barred. We have checked the bar fractions as a function of bar length and
conclude that Seyferts have more bars at practically all lengthscales or, at
most, the bar fractions are equal. We estimate a formal significance of the
result that Seyfert host galaxies are barred more often than non-active
galaxies at the 2.5$\sigma$ level, based on  the quadratic combination of the
uncertainties for the individual samples.  Our current results are in perfect
agreement with those of KSP (Table~1) but have a higher significance. 

The overall bar fraction of 62\% $\pm$ 5\% compares with the result of 69\%
$\pm$ 6\% as found from the NIR imaging analysis of our combined Seyfert and
control samples in KSP, and with other determinations in the literature,
ranging from below 60\% to around 75\% (e.g., Mulchaey \& Regan 1997; Eskridge
et al. 2000). Whereas NIR imaging surveys have led to a factor of two increase
in the bar fraction with respect to the RC3 (SB classification), the bar
fractions determined in the current work are slightly lower. This is clearly
related to our choice of a conservative approach and strict criteria for bar
identification (in fact, slightly more restrictive than in KSP), and our aim
for sample comparison, rather than establishing absolute numbers. Use of
subjective and non-reproducible criteria, as, e.g., in Eskridge et al. (2000)
and in all major galaxy catalogues, may well lead to higher bar fractions.

\subsection{Bar Strength in Seyfert and Non-Seyfert Galaxies}

We count all the bars with ellipticities $>$ 0.45 as ``strong'' bars, although
the axial ratio does not necessarily reflect the strength of a bar which is
measured by the maximal ratio of tangential-to-radial gravitational forces
(Shlosman, Peletier, \& Knapen 2000). This division gives 55\% $\pm$ 7\% strong
bars of all sizes among Seyferts and 57\% $\pm$ 8\% among non-Seyferts. The
same procedure applied to the secondary bars, using the same division between
strong and weak bars, results in 29\% $\pm$ 12\% strong secondary bars among
Seyfert galaxies, and 43\% $\pm$ 19\% among non-Seyferts. Figure 2 contrasts the
ellipticities (divided into two groups, separated by 0.45), between primary and
secondary bars, and nested and single  bars, both separately for Seyferts and
non-Seyferts. It also compares all Seyfert bar ellipticities to all non-Seyfert
bar ellipticities. Significantly, Seyferts always have more or at least the
same number of weak bars as non-Seyferts, among secondary, primary and single
bars (Fig.~2). This reinforces the earlier results of Shlosman  et al. 
that the bars in Seyferts are weaker than the bars in non-Seyferts. 

\subsection{Comparison Between Seyfert 1 and Seyfert 2 Galaxies}

Because the unified theory of AGNs (e.g. Antonucci 1993) states that Seyfert 1
and Seyfert 2 galaxies are intrinsically similar, we study the bar fractions 
separately for Seyfert 1 and 2 galaxies. We group Seyfert 1--1.9 galaxies
together as Seyfert 1 and compare their properties to the Seyfert 2  class
(which included the Sy1h galaxies where the broad lines are seen in polarized
light) and find that 16/23 (70\% $\pm$ 10\%) of Seyfert~1 galaxies  possess at
least one bar. In comparison, 25/33 (76\% $\pm$ 7\%) of Seyfert~2 galaxies
have at least one bar. Therefore, the bar fractions are the same within
uncertainties among the two Seyfert types. 

\includegraphics[width=3.0in]{fig6cor.eps}
\figcaption{Normalized bar lengths for all the bars in the Seyfert sample.
Seyfert~1 bars are shown with filled circles and Seyfert 2 bars with open
circles. The  x-axis is the  normalized bar length after deprojection,
(normalized to the galaxy diameter, tabulated in Tables 6 and 7) and the y-axis
is the ellipticity after a  two-dimensional  deprojection.\label{fig6}}
\vskip0.1in

We found that 13\% $\pm$ 7\% of the Seyfert 1 galaxies and 27\% $\pm$ 8\% of
Seyfert 2 galaxies have nested bars. This difference could be due to the more
luminous nuclei of Seyfert 1 galaxies hiding secondary bars or, in fact,
may represent a more fundamental property of these galaxies, to be decided
in larger samples. Lastly, Figure~6 shows that distributions of the normalized
lengths of the Seyfert bars (normalized by the galaxy diameter) versus
ellipticity separately for Seyfert 1 and Seyfert 2 galaxies are similar. 

\section{DISCUSSION}

\subsection{Nested and Single Bar Properties}

\subsubsection{Length Separation Between Primary and Secondary Bars}

The observed nested bars in our samples of Seyfert and non-Seyfert  galaxies
reveal an intriguing property --- the existence of a critical physical length,
$\approx$ 1.6~kpc, which separates the primary and the secondary bars,
resulting in a clear bimodal size distribution with only little overlap.
Moreover, when the bar sizes are normalized to those of the respective host
galaxies, the overlap between the two bar species is further reduced. To
illustrate this effect, we have constructed Figure~7, where the physical sizes
of deprojected bars are shown vs. $D_{25}$.  We note that in this figure the
primary  bar sizes exhibit a roughly linear correlation with the parent galaxy
sizes (the linear correlation coefficient is 0.66 and the probability that this
is achieved by uncorrelated points is less than 1\%). The slope of this
correlation is finite and non-zero. On the other hand, the secondary bar sizes
are limited by an upper boundary. This can be interpreted as a linear
correlation with a zero slope or in other words, the sizes of these bars are
{\it independent} from the sizes of their host galaxies.  The importance of
this result can be inferred from the fact that only in this case the normalized
(to $D_{25}$) bar lengths will preserve the identity of both bar groups and
there will be no further mixing between the primary and secondary bars in the
normalized size space, as shown in Figs.~3 and 4. If, for example, both types
had a linear correlation with $D_{25}$ where both slopes were non-zero, the two
bar groups (primary and secondary bars) would be separated in physical space,
but mixed in the normalized space.

\vskip0.1in

\includegraphics[width=3.0in]{fig7cor.eps}
\figcaption{Primary (open squares) and secondary (asteriscs) bar sizes vs
$D_{25}$. \label{fig7}} 
\vskip0.1in

We suggest a simple and attractive explanation for the presence
of a bimodal distribution of bar sizes both in the physical and in the
normalized space, related to the property that secondary bars are limited to
within $\sim 1.6$~kpc in their physical size. The linear correlation of the
primary bar size with galaxy size means that these bars extend to a fixed
number of radial scalelengths in the disk. The absence of such a correlation
for the secondary bars, together with their limited range of sizes, hints to a
different physical nature of formation and dynamics compared to the primary
bars. 

As discussed in Section~2, numerical simulations of nested bars show that the
secondary bars are confined to the region within the ILRs of the primary bars.
These ILRs develop close to the radius of the rotation velocity  turnover (or a
substantial bump in the rotation curve), which is generally  located just
outside the bulge (at least for early Hubble types), or more basically, where
the mass distribution in the inner galaxy switches from 3D to 2D with
increasing radius. In a  hypothetical case of a plane-parallel and uniform
galactic disk, this happens at $r\sim \Delta z$, where $\Delta z$ is the
thickness of the disk. For a realistic surface density distribution in the
disk and in the presence of a bulge, one can use $\sim$ 1--2~kpc as a
reasonable estimate for the position of the ILRs (in fact of the outer ILR). 
For the early Hubble type galaxies (S0-Sb), this leads to the appearance of an
ILR at about the bulge radius. For the later Hubble type galaxies  or early
types with small bulges, the height of the disk becomes comparable to the
radius of the disk at a point where the 2D disk approximation breaks down. 
Further evidence for a constant normalized position of the ILR in disk
galaxies of all sizes is provided by Athanassoula \& Martinet (1980) and
Martin (1995), who found a linear correlation between the large-scale bar and
bulge sizes. 

If primary and secondary bars had a similar formation and evolution history,
one would expect a linear correlation between the secondary bar length and
$D_{25}$, simply because of the observed correlations between the disk,
large-scale bar, and bulge sizes. However this correlation is clearly ruled out
by our data. A possible resolution of this discrepancy is as follows.

Large-scale bars are known to extend to about $(0.83\pm 0.12)$ of their
corotation radii, an empirical rule based on the shapes of their offset dust
lanes (Athanassoula 1992). However, the {\it secondary} bars are not expected
to follow this rule or to possess offset dust lanes (Shlosman \&
Heller 2001). Because a large degree of dissipation is involved in forming
these small bars, their size can be much smaller than their corotation radius
(which is also the ILR of the primary bar). This particular property of
secondary bars is expected to destroy any correlation between their size and
that of the parent galaxy.

We have compiled a sample of 62 galaxies with nuclear rings (mostly from Buta
\& Crocker 1993) and determined their normalized size distribution (see
Figure~8). This distribution peaks at $r_{\rm ring}/D_{25}$ = 0.06, supporting
our view that it acts as the  dynamical separator between secondary and
primary bars. This result is consistent with our claim that the secondary bar
lengths are limited by the size of the ILR, since nuclear rings are associated
with  the ILRs (e.g., Schwarz 1984; Combes \& Gerin 1985; Knapen et al.
1995b). 

\subsection{Comparison with Earlier Studies of Secondary Bars}

\subsubsection{Ground-Based Studies}   

Previous ground-based studies (e.g. Shaw et al. 1995; Jungwiert et al. 1997;
Erwin \& Sparke 1999a) have found secondary bars in 20\%--25\% of their sample
galaxies. Only a fraction of their secondary bars fulfills our bar criteria. In
addition, there are biases in these samples, e.g., Shaw et al. (1995) and Erwin
\& Sparke (1999a) start with samples where the outer bar is well detected. Shaw
et al. (1995) define as secondary bars those NIR isophote twists which occur
inside the minor axis width of the large-scale bar and where the secondary bar
is not aligned with the large-scale bar. They found at most seven (24\%) such
cases in a sample of 29 relatively face-on galaxies. Ellipticity and position
angle profiles were given only for five of these objects. Of these only one,
NGC~4321, fulfills our criteria for a bar.

\vskip0.1in

\includegraphics[width=3.0in]{f8.eps}
\figcaption{The number distribution of normalized nuclear ring diameters.  Data
were mostly taken from Buta \& Crocker (1993), but we added a few  ``famous''
nuclear ring galaxies.\label{fig8}} 
\vskip0.1in  

Jungwiert et al. (1997) considered a sample of 56 galaxies with inclinations
less than 75$\degr$. We note that it may be exceedingly difficult to find bars
at  such high inclinations. Jungwiert et al. have found 17 galaxies (30\%) with
two triaxial (and therefore, possibly double-barred) structures. Of these, only
8 are secondary bars according to our criteria. In addition, they find bar
signatures (ellipticity peaks with constant position angles) in the nuclear
regions (less than 1 kpc) of three SA galaxies. Deprojection effects were also
studied and accounted for in a simple two-dimensional approximation. Erwin \&
Sparke (1999a) find 5 (23\%) definitely double barred galaxies in a sample of
22 barred galaxies by fitting ellipses to the isophotes. Only two profiles are
given and of these only one fits our bar criteria. In addition, they find
another five possible double-barred systems. One of their definitely barred
galaxies, NGC ~2681, is claimed to be triple-barred.

\subsubsection{Previous {\it HST} Studies of Small Bars}

Recent papers by Regan \& Mulchaey (1999) and Martini \& Pogge (1999) suggest
that nuclear bars may not be important for fueling AGN activity because nuclear
bars were found in only a small minority of their samples. Specifically, Regan 
\& Mulchaey (1999) claim that the Seyferts Mrk 573 and Mrk 1066 do not have
nuclear bars, based on their nuclear dust morphology. However, we have found
sub-kpc bars in both of these galaxies. On the other hand, Regan \& Mulchaey
claim that NGC 5347 and NGC 7743 have nuclear bars, whereas we have not
detected them by ellipse-fitting. Furthermore, Martini et al. (2001)
discuss the role of secondary bars in the fueling of the nuclear activity in
disk galaxies without invoking a matched control sample of non-active galaxies.

One should be aware of the caveats associated with the Regan \& Mulchaey (1999)
result. First, they looked for straight segments of dust lanes in the
circumnuclear region. Such dust lanes are usually seen on the leading side of
large-scale bars. However, the physical conditions in the circumnuclear region
are known to differ from those at kpc scales, and there is no reason why gas
and stellar dynamics  should be identical as well. Shlosman \& Heller (2001)
have analyzed the gas flow in secondary bars, and found that gas flow in the
secondary bars differs from that in the primary bars due to a time-dependent
potential, fast rotation, specifics of gas crossing the bar--bar interface, and
other reasons related to secondary bar formation. No offset dust lanes form
under these conditions.

\subsection{Excess of Bars among Seyfert Galaxies}

\subsubsection{Comments on Bar Detection}

The main result from the comparison of Seyfert and non-Seyfert galaxy bars that
emerges from our study is that Seyfert galaxies have more bars on almost any
lengthscale. Evidently on the largest scales stellar bars (spontaneous and
induced) are so frequent that when taken together with oval distortions of
disks, they become nearly universal. Under these circumstances it is clear
that  large-scale bars are an important but not a sufficient factor to fuel the
nuclear activity in AGNs, as was already pointed out by Shlosman et al. (1989).
It has been established beyond doubt during the last decade that large-scale 
bars are efficiently channeling gas towards the central kpc and induce
starburst activity there, mainly in the form of nuclear rings. But this does
not explain the fate of the inflowing gas at even smaller scales, between a few
100 pc and 1~pc. The mere existence of secondary bars on these small spatial
scales hints that gravitational torques are important here, but the exact role
of secondary bars in the fueling hierarchy, as well as the details of their
formation and evolution are obscure. Here we focus on the properties of stellar
bars in the NIR, but one should not forget that at least one additional factor
must play a crucial role in differentiating between Seyfert and non-Seyfert
hosts, the availability of digestable fuel, and therefore the knowledge of
dynamics of the self-gravitating gas in the background potential of nested
bars is of prime importance (Shlosman et al.).

Given the poor record of detecting large-scale bars in the optical, the bulge
light dilution of secondary bar isophotes in the NIR, effects of dust and
star formation within the central kpc (even in the NIR), and finally our
conservative approach to bar detection, it is not surprising that in this paper
we have detected only modest bar fractions in disk galaxies. We have found that
stellar dynamics cannot be reliably traced even by the use of NIR
photometry. This should be taken into account while assessing the significance
of  recent studies in this field, such as the work of Martini et al. (2001).
Nevertheless, this fraction by far exceeds the fraction of galaxies classified
as ``SB'' in any optical catalogue, e.g., the RC3. 

It is known, for example, that Sy~2 galaxies, which comprise the large majority
of our Seyfert sample, are dustier and include more star formation (e.g.,
Gonz\'alez Delgado, Heckman, \& Leitherer 2001). The deprojection procedure
will introduce additional uncertainties in the bar axial ratios. On the other
hand, there is probably not enough dust present even in the inner regions of
spirals to hide many bars in the $H$-band. Giovanelli et al. (1994) showed that
in the Cousins $I$-band the central optical depth is smaller than 5. Assuming
the Galactic extinction law (Rieke \& Lebofsky 1985), this corresponds to an
optical depth of 1.6 in $H$. Since this value goes down rapidly with decreasing
inclination, most galaxies have central optical depths in $H$ smaller than 1
mag (in agreement with Peletier et al. 1995), implying that it is hard to hide
bars in the circumnuclear regions of disk galaxies. Possibly the largest
unknown in the overall picture of nested bars is the lifetime of secondary bars
which will affect directly their observed frequency.  

One should also remark on the deprojected ellipticity distribution of bars on
all spatial scales. We find that the ellipticity distribution peaks at about
$\epsilon \sim 0.4-0.5$. The numbers decline for larger ellipticities, as well
as for smaller ellipticities. The same trend is preserved if only relatively
face-on galaxies are counted, say with axial ratios $>$0.85. No theoretical
explanation exists presently for this behavior.

\subsubsection{Galaxy Interactions: Induced and Spontaneous Bars in
Disk Galaxies}

The effects of interactions on inducing the formation of stellar bars have not
been exhaustively studied. Noguchi (1988) and Salo (1991) suggested that flybys
induce stellar bars when tidal forces exceed about 10\% of the unperturbed
radial force and when the encounter is prograde. There is no indication so far
that the gas flow pattern differs between induced and spontaneous (i.e., those
formed as a result of bar instability) bars. Because such tidally-induced bars
are indistinguishable from bars formed in a bar instability, and because we aim
at a detailed comparison between the occurrence and properties of bars in
Seyfert and non-Seyfert control galaxies, irrespective of their origin, our
strategy is not to exclude interacting galaxies, apart from the strongly
distorted ones. In other words, for the purpose of understanding the
morphological differences which may be responsible for fueling the central
activity in disk galaxies, it is irrelevant which type of bar is hosted by the
galaxy. One needs only to exclude the strongly interacting galaxies which are
heavily distorted and, therefore, complicate the bar identification. 

\begin{table*}
\caption{Bar fractions and presence of faint or bright companions, for
Seyfert and control samples.\label{tab2}}
\begin{tabular}{lccccc}
\tableline
\tableline
Sample  & \multicolumn{2}{c}{Bar fraction: Seyferts} &
\multicolumn{2}{c}{Bar fraction: control}\\
& $N$ & Percentage & $N$ & Percentage \\
\tableline
Overall              & 41/56 & 73\% $\pm$  6\% & 28/56 & 50\% $\pm$ 7\%\\
No companions\tablenotemark{a}    & 16/22 & 73\% $\pm$ 9\% &  5/10 & 50\% $\pm$ 16\%\\
Companion(s)\tablenotemark{a}     & 25/34 & 74\% $\pm$  8\% & 23/46 & 50\% $\pm$  7\%\\
Not interacting\tablenotemark{b} & 24/33 & 73\% $\pm$  8\% & 16/33 & 48\% $\pm$  9\%\\
Interacting\tablenotemark{b}     & 17/23 & 74\% $\pm$  9\% & 13/23 & 57\% $\pm$ 10\%\\
\hline
\end{tabular}
\tablenotetext{a}{Indicates presence or absence of
companion galaxies within 400 kpc in radius and within $\pm500$ km~s$^{-1}$
in $cz$.}
\tablenotetext{b}{Idem, but additionally companion galaxy may not be fainter
by $B_T=1.5$ mag than the sample galaxy under consideration for the
latter to be qualified as ``interacting'' (see text).}
\end{table*}  

The criticism by M\'arquez et al. (2000) of the KSP results by implying that
the  excess of bars among Seyfert hosts is due to the inclusion of galaxies
that may be undergoing gravitational interaction is therefore not
justified. Moreover, it is unclear what can be achieved by following the
M\'arquez et al. procedure, which reduced the KSP sample to 13 (Sy) and 11
(control) galaxies by rejecting all galaxies that have companions within a
cylindrical volume of 0.4~Mpc in projected radius and 2$\times$500~km~s$^{-1}$ 
in $cz$ (distance along the line of sight, corresponding to 6.7~Mpc when
assuming a Hubble flow with $H_0$=75~km~s$^{-1}$/Mpc). Clearly, the few
remaining galaxies in the sample are insufficient to allow any meaningful
conclusions on barred fractions  in different samples. The derived numbers are
not  statistically significant and cannot be interpreted as evidence against
higher bar fractions among Seyferts.  

We note that although Seyfert activity occurs in interacting and merging
galaxies, there is no significant evidence for an excess of companions to
Seyfert galaxies as compared to non-active control galaxies (Fuentes-Williams
\& Stocke 1988; de Robertis, Yee, \& Hayhoe 1998, etc.). Earlier work by e.g. 
Adams (1977) and Dahari (1984) was plagued by poor control sample selection.
This implies that there is no a priori reason to reduce samples of Seyfert and
non-active galaxies artificially by excluding all galaxies with companions; the
only effect of such an operation is the  reduction of the numbers of galaxies
in both samples in equal amounts, which effectively corresponds to an increase
in the statistical uncertainty of the  final result.

In order to check the statements above with the galaxies in our samples, we
have used the Lyon-Meudon extragalactic database (LEDA) to find companions to
all our sample galaxies. As a first step we used the same criteria as M\'arquez
et al. (2000) to find those sample galaxies which have companions within 400
kpc in radius, and within $\pm500$ km~s$^{-1}$ in $cz$. We found that 34 (out
of 56) or 61\% $\pm$ 7\% of our Seyfert and 46 (out of 56) or 82\% $\pm$ 5\% of
our control galaxies in fact have companions within such a volume. We also
found, though, that this fraction reaches almost 100\% for the nearest galaxies
in our sample, in line with experience that most if not all galaxies will have
companions of some size (e.g., the Milky Way, M31).

We refined our search in the second step, where we impose the additional
criterion that companion galaxies may not be fainter than the sample galaxy by
$\Delta B_T=1.5$ mag. This limit is somewhat arbitrary but ensures that the
companion to M51, NGC~5195, is included under these criteria. We call the
sample galaxies with such bright companions ``interacting'' although they may
not be so at a level which distorts their appearance. We find that in both
the Seyfert and the control sample, 23 out of 56 (41\% $\pm$ 7\%) of sample
galaxies are interacting, while the remaining 33 (59\% $\pm$ 7\%) are not. We
thus find no evidence for a difference between the Seyfert and control sample
in terms of bright companions, in agreement with the recent findings of Schmitt
(2001), but do find that our control galaxies have (faint) companions
significantly more often than the Seyferts do. This cannot be an effect of
closer distance of the control galaxies since we matched them in distance to
the Seyfert sample.

The important question in relation to the current paper is whether the presence
of these companions, bright or faint, is related to the presence of a bar (as
claimed by M\'arquez et al. 2000). The bar fractions, and associated
uncertainties from Poisson statistics, are given in Table~2 for all the
subsamples as defined above (Seyfert and control, with and without faint and
bright companions). The conclusion from these numbers is that the bar fraction
among our sample galaxies is completely independent of the presence of faint,
or bright, companions. This conclusion does not contradict the earlier study by
Elmegreen, Elmegreen, \& Bellin (1990), which showed that there is an excess of
bars among early-type galaxies in strongly interacting pairs. Elmegreen et al.
used the ``SB'' classification in optical catalogs instead of ellipse fitting
to find bars, and their sample consisted of much closer galaxy pairs. In
summary, we have addressed the kind of criticism raised by M\'arquez et al. by
demonstrating that the presence of a bar in our sample galaxies is not related
to the presence of companions.

\section{CONCLUSIONS}

We have taken advantage of the high spatial resolution offered by the  {\it
HST} and the large database of disk galaxies of Hubble types S0--Sc in the {\it
HST} archive, complemented by ground-based NIR and optical images of the outer
disks, to examine the numbers and properties of bars in 56 Seyfert and 56
matched non-Seyfert galaxies. We emphasize that our results still suffer from 
small number statistics and, therefore, must be confirmed by a study of larger
samples. The use of adaptive optics on ground-based telescopes and further
imaging by the {\it HST} promises to increase the number of suitable Seyfert
and non-Seyfert galaxies substantially in the near future. This will allow the
determination of the exact dynamical role that nested bars play in the
evolution of disk galaxies and the fueling of central activity, stellar and
nonstellar. 

Our results are not sensitive to the exact definition of a bar. To  verify
this, we have varied the cutoff ellipticity of the bar within 20\% and the
range of allowed variation in the bar position angle within 50\%.
Those  resulted in insignificant changes in the bar statistics. Our main
results are  listed in the following.

\begin{enumerate}

\item We find that primary and secondary bar sizes, both physical and
normalized by the galaxy diameter $D_{25}$, show a bimodal distribution with
little overlap between the two groups. The separating value, when normalized by
the galaxy diameter $D_{25}$, is around 0.06 (i.e., 0.12 of the corresponding
galaxy radius). In a physical space, the dividing length between the primary
and secondary bars is about 1.6~kpc. We identify these critical values with the
location of the inner Lindblad resonances (ILRs) --- dynamical separators
between the nested bars. The ILRs are expected to form where the gravitational
potential of the inner galaxy switches from 3D to 2D. This happens at the
bulge--disk interface, or alternatively, where the disk thickness becomes
comparable to its radius.

\item The distribution of nuclear ring sizes (radii) obtained from the
literature was found to peak at the same normalized critical length of 0.06. A
nuclear ring is considered to be a clear indicator of the inner Lindblad
resonance (ILR) in the disk and is formed just interior to this resonance in
all numerical simulations of gas flows in barred galaxies. We interpret the
correlation between the nuclear ring sizes and the critical value of 0.06 in
nested bars as an additional strong indication in favor of the crucial roles
the ILRs play in the dynamics of these systems.   

\item Primary bars in nested systems show a linear correlation with the size of
the host galaxy disk, extending to a fixed number of disk scalelengths.

\item We find that the secondary bar sizes do not correlate with the disk
sizes, primary bar sizes, nor with the critical size of 0.06. As a corollary,
these bars do not correlate with the radii of the primary ILRs (which coincide
with the secondary bar corotation radii). This is contrary to the behavior of
primary (and single) bars which are known to extend to within $0.83\pm 0.12$ of
their corotation radii. 

\item Within the nested bars, the secondary bars have smaller ellipticities than the 
primary bars. This can be simply explained by noting that the secondary
bars lie within the galactic bulges which dilute the observed small bar
ellipticities, making them rounder.                                         

\item Both the single and primary bars show a correlation between the
ellipticity and the length of the bar. This result has been reported earlier
for late-type disk galaxies, but it is extended here for galaxies which range
in Hubble type from S0 to Sc.    

\item A relatively small fraction of {\it single} bars are shorter than the 
critical value of 0.06, the boundary between the primary and secondary bars
in nested systems. Single bars also have higher average ellipticities than
nested bars, and a different distribution in morphological types. Although the
numbers are small, this raises the interesting possibility that single bars
have a different formation mechanism.

\item The comparison of bar numbers between Seyfert and non-Seyfert galaxies
shows that Seyfert galaxies have an excess of bars, namely 73\% $\pm$ 6\% of
Seyferts have at least one bar, against only 50\% $\pm$ 7\% of non-Seyfert
galaxies. The statistical significance of this result is at the 2.5$\sigma$
level, and confirms and strengthens the result of KSP which was based on
smaller samples.

\item We confirm numerically that within our samples the presence of
companions, even bright ones, near a galaxy bears no relation to the presence
of a bar in that galaxy.

\item Seyfert galaxies have thicker (in the sense of axial ratio $b/a$) bars on
average than the non-Seyfert galaxies, no matter how the comparison is made
(among primary bars, secondary bars, single bars, or as a function of the host
galaxy Hubble type). We thus confirm our earlier result of a deficiency of thin
bars among Seyferts in a study of the CfA sample of Seyferts and a matched
control sample of non-Seyfert galaxies.

\item We find a difference in the fraction of nested bars among Seyfert
1 galaxies (13\% $\pm$ 7\%) and Seyfert 2 galaxies (27\% $\pm$ 8\%). This
effect can be most probably explained by the very luminous Seyfert 1 nuclei.
\end{enumerate}                                

Overall, we find that NIR isophote fitting, a highly reliable method of
detecting large-scale stellar bars, shows difficulties when applied to sub-kpc
bars. The main difficulty comes from localized and distributed sites of dust
extinction and bright stars within the central kpc. This results in a
substantial underestimate of bar fraction.

\acknowledgments

We are grateful to Lia Athanassoula, Albert Bosma, Tim Heckman and Clayton
Heller for illuminating discussions. We thank Fran\c{c}oise Combes, Zsolt Frei,
Isabel M\'arquez, and Kartik Sheth for providing us with some images. SL thanks
Richard Hook for advice on the use of the CPLUCY deconvolution routine and S.
Jogee for bringing into our attention a flaw in the deprojection program which
we subsequently corrected. Our work is based on observations with the NASA/ESA
{\it HST}, obtained from the data archive at the STScI, which is operated by
the Association of Universities for Research in Astronomy, Inc. under NASA
contract No. NAS5-26555. Our research is partly based on  observations made
with ESO telescopes at La Silla. The J. Kapteyn and W. Herschel Telescopes are
operated on the island of La Palma by the Isaac Newton Group (ING) in the
Spanish Observatorio del Roque de los Muchachos of the IAC. Data were retrieved
from the ING archive. This research has made use of the NASA/IPAC Extragalactic
Database (NED) which is operated by the JPL, Caltech, under contract with NASA.
We have made use of the LEDA database (http://leda.univ-lyon1.fr). The DSSs
were produced at the STScI under U.S. Government grant NAG W-2166. The images
of these surveys are based on photographic data obtained using the Oschin
Schmidt Telescope on Palomar Mountain and the UK Schmidt Telescope. This
publication makes use of data products from the 2MASS, which is a joint project
of the University of Massachusetts and IPAC/Caltech, funded by NASA and NSF. IS
acknowledges support from NASA grants HST AR-07982.01-96A, GO-08123.01-97A, NAG
5-3841, NAG 5-10823  and WKU-522762-98-6, and thanks the organizers of INAOE
workshop on {\it Disk Galaxies: Kinematics, Dynamics and Perturbations} for
supporting a prolonged visit during which this work was finalized.

\appendix

\section{GRAPHICAL AND TABULAR REPRESENTATION OF SAMPLE MATCHING}

The details of how we matched the Seyfert galaxy sample properties to those of
a control sample of non-Seyfert galaxies were explained in Section 3.2. In this
Appendix we show a tabular and graphical representation of the galaxies in  the
various bins of the four quantities that were matched, distance, absolute $B$
magnitude, axial ratio, and morphological type, in Table~A1 and Figure~A1,
respectively. The general properties of the galaxies are given in Tables~A2 and
A3.

\setcounter{figure}{0}
\begin{figure*}[hp]
\centering
\includegraphics[width=6.0in]{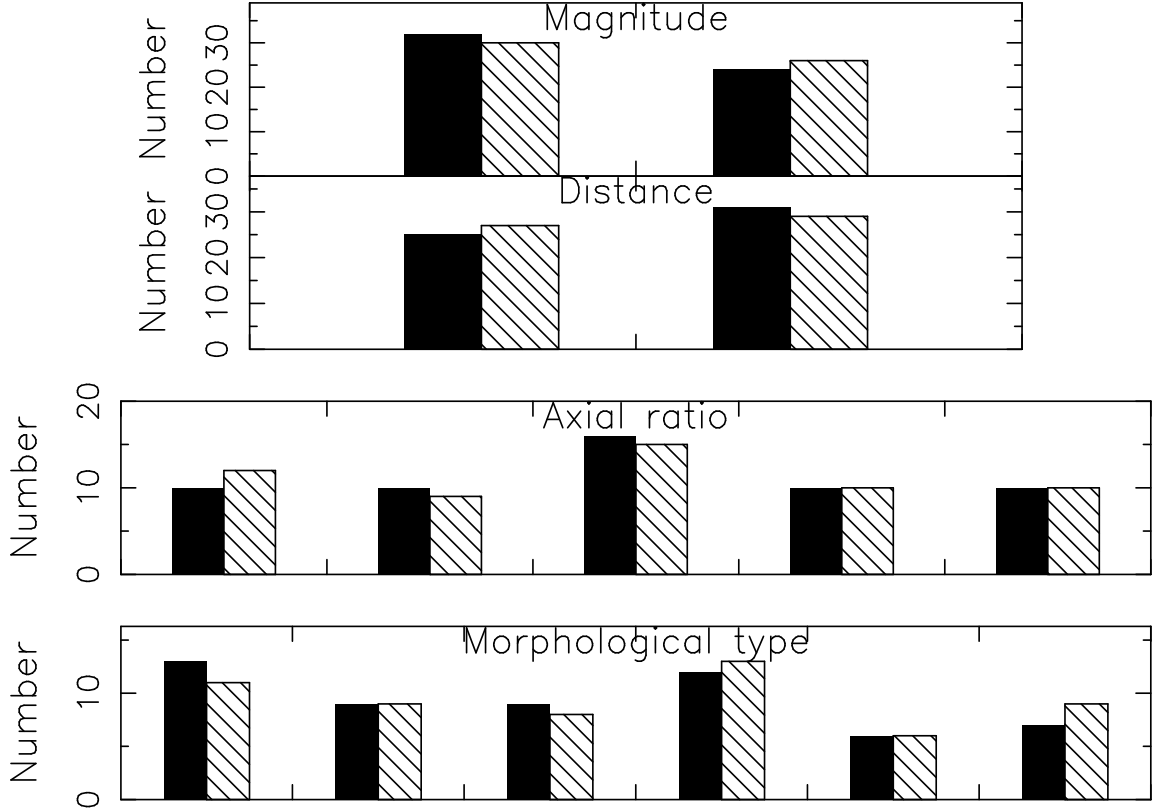}
\figcaption{Matching of the Seyfert and non-Seyfert samples with respect to
four different properties. The absolute $B$-magnitude separator is -20.4 mag.
The distance separator is 38 Mpc. The axial ratio bins are separated by
0.6, 0.7, 0.8, and 0.9. Finally, the morphological T-types are separated by
0.1, 1.1, 2.1, 3.1, and 4.1. The borders between the bins were selected so
that we have approximately equal numbers in each bin. The Seyfert galaxies are
shown with filled columns and non-Seyferts with hatched columns.\label{fig9}}
\end{figure*}

\setcounter{table}{0}
\begin{deluxetable}{lccccccccccccccc}
\tabletypesize{\scriptsize}
\tablecaption{Match of the Seyfert and non-Seyfert galaxy samples
 with respect to absolute $B$ magnitude, distance, axial ratio and
 morphological type. \label{tab3}}
\tablewidth{0pt}
\tablehead{
\colhead{Type of galaxy} & \colhead{Mag$_{1}$} & 
\colhead{Mag$_{2}$} & \colhead{Dist$_{1}$} & \colhead{Dist$_{2}$} & 
\colhead{{\it b/a}$_{1}$} & \colhead{{\it b/a}$_{2}$} & 
\colhead{{\it b/a}$_{3}$} & \colhead{{\it b/a}$_{4}$} & 
\colhead{{\it b/a}$_{5}$} & \colhead{T$_{1}$} & \colhead{T$_{2}$} & 
\colhead{T$_{3}$} & \colhead{T$_{4}$} & \colhead{T$_{5}$} & \colhead{T$_{6}$}    
}
\startdata 
Seyferts & 32 & 24 & 25 & 31 & 10 & 10 & 16 & 10 & 10 & 13 & 9 & 9 & 12 & 6 & 7
\\ 
Non-Seyferts & 30 & 26 & 27 & 29 & 12 & 9 & 15 & 10 & 10 & 11 & 9 & 8 & 13 & 6 &
9 \\
\enddata 

\tablecomments{Number of galaxies with absolute $B$ magnitude $<$ -20.4 (col.
2); Number of galaxies with absolute $B$ magnitude $\geq$ -20.4 (col. 3);
number of galaxies with distance smaller than 28 Mpc (col. 4); number of
galaxies with distance larger than 28 Mpc (col. 5); number of galaxies with
$b/a$ $<$ 0.6, 0.6 $\leq$ $b/a$ $<$ 0.7, 0.7 $\leq$ $b/a$ $<$ 0.8, 0.8 $\leq$
$b/a$ $<$ 0.9, and b/a $\geq$ 0.9 (cols. 6, 7, 8, 9, and 10, respectively); number of
galaxies with morphological T-types T $<$ 0.1, 0.1 $\leq$ T $<$ 1.1, 1.1 $\leq$
T $<$ 2.1, 2.1 $\leq$ T $<$ 3.1, 3.1 $\leq$ T $<$ 4.1, and T $\geq$ 4.1 (cols.
11, 12, 13, 14, 15, 16, respectively).}

\end{deluxetable}

\begin{deluxetable}{lccccccccc}
\tabletypesize{\scriptsize}
\tablecaption{Properties of Seyfert galaxy sample. \label{tab4}}
\tablewidth{0pt}
\tablehead{
\colhead{Galaxy} & \colhead{{\it HST} source} & \colhead{Other source} & \colhead{Class.} & 
\colhead{T} & \colhead{Sy} & \colhead{{\it b/a}} & \colhead{V$_{hel}$} & 
\colhead{D} & \colhead{M$_{B}$}   
}
\startdata 
ESO 137-G34 & Mu & DSS (R) & SAB(s)0/a? & 0.0 & Sy2 (Veron) & 0.76 & 2747 & 36.6 & -23.47 \\
IC 2560 & Mu & DSS (R) & (R':)SB(r)bc & 3.3 & Sy2 (Veron) & 0.63 & 2925 & 39.0 & -21.04 \\
IC 5063 & Mu & ESO (V), DSS (R) & SA(s)0+: & -0.8 & Sy1.9 (Veron) & 0.67 & 3402 & 45.4 & -20.71 \\
Mrk 573 & Mu, Po & DSS (R) & (R)SAB(rs)0+: & -1.0 & Sy2 (Veron) & 1.00 & 5174 & 69.0 & -20.62 \\
Mrk 1066 & Mu & DSS (R) & (R)SB(s)0+ & -1.0 & Sy2 (Veron) & 0.59 & 3605 & 48.1 & -20.37 \\
Mrk 1210 & Mu & DSS (R) & Sa & 1.0 & Sy1h (Veron) & 1.00 & 4046 & 53.9 & -19.45 \\
NGC 788 & Mu & DSS (R) & SA(s)0/a: & 0.0 & Sy1h (Veron) & 0.74 & 4078 & 54.4 & -20.88 \\
NGC 1068 & Th & 2M (H), DSS (R) & (R)SA(rs)b & 3.0 & Sy1h (Veron) & 0.85 & 1137 & 14.4 & -21.32 \\
NGC 1241 & Mu & 2M (H), DSS (R) & SB(rs)b & 3.0 & Sy2 (Veron) & 0.61 & 4062 & 26.6 & -20.04 \\
NGC 1365 & St & ESO (R), DSS (R) & (R')SBb(s)b & 3.0 & Sy1.5 (Veron) & 0.55 & 1636 & 16.9 & -21.21 \\
NGC 1667 & Mu & 2M (H), DSS (R) & SAB(r)c & 5.0 & Sy2 (Veron) & 0.78 & 4547 & 60.6 & -21.50 \\
NGC 1672 & Mu & DSS (R) & (R'{\_}1:)SB(r)bc & 3.0 & Sy2 (Veron82) & 0.83 & 1350 & 14.5 & -20.56 \\
NGC 2639 & Mu & 2M (H), DSS (R) & (R)SA(r)a:? & 1.0 & Sy1.9 (Veron) & 0.61 & 3336 & 44.5 & -21.05 \\
NGC 2985 & Mu, St & 2M (H), DSS (R) & (R')SA(rs)ab & 2.0 & Sy1.9(Veron) & 0.79 & 1322 & 22.4 & -20.77 \\
NGC 3031 & St & 2M (H), ING (I), DSS (R) & SA(s)ab & 2.0 & Sy1.5 (Veron) & 0.52 & -34 & 1.4 & -18.34 \\
NGC 3081 & Mu & 2M (H), DSS (R) & (R{\_}1)SAB(r)0/a & 0.0 & Sy2 (Veron) & 0.78 & 2385 & 32.5 & -19.97 \\
NGC 3227 & Ri & 2M (H), DSS (R) & SAB(s) pec & 1.0 & Sy1.5 (Veron) & 0.67 & 1157 & 20.6 & -20.39 \\
NGC 3486 & Mu & 2M (H), DSS (R) & SAB(r)c & 5.0 & Sy2 (Ho) & 0.74 & 681 & 7.4 & -18.58 \\
NGC 3516 & Mu & 2M (H), DSS (R) & (R)SB(s)$\hat{}$0$\hat{}$0: & -2.0 & Sy1.5 (Veron) & 0.78 & 2649 & 38.9 & -20.81 \\
NGC 3718 & Mu & 2M (H), DSS (R) & SB(s)a pec & 1.0 & Sy1 (Veron) & 0.49 & 994 & 17.0 & -19.96 \\
NGC 3786 & Po & 2M (H), DSS (R) & SAB(rs)a pec & 1.0 & Sy1.8 (GR) & 0.59 & 2678 & 41.6 & -18.18 \\
NGC 3982 & Mu, Po & CF (H), DSS (R) & SAB(r)b: & 3.0 & Sy1.9 (Veron) & 0.87 & 1109 & 17.0 & -19.47 \\
NGC 4117 & Mu & 2M (H), DSS (R) & S0$\hat{}$0$\hat{}$: & -2.3 & Sy2 (Veron) & 0.49 & 958 & 17.0 & -17.12 \\
NGC 4151 & Th & CF (H), 2M (H), DSS (R) & (R')SAB(rs)ab: & 2.0 & Sy1.5 (Veron) & 0.71 & 995 & 20.3 & -20.83 \\
NGC 4253 & Mu & DSS (R) & (R')SB(s)a: & 1.0 & Sy1.5 (Veron) & 0.80 & 3876 & 56.7 & -19.98 \\
NGC 4303 & Mu & DSS (R) & SAB(rs)bc & 4.0 & Sy2 (Veron) & 0.89 & 1566 & 15.2 & -20.79 \\
NGC 4593 & Mu & 2M (H), DSS (R) & (R)SB(rs)b & 3.0 & Sy1 (Veron) & 0.74 & 2698 & 39.5 & -21.55 \\
NGC 4725 & Mu & Frei (I), DSS (R) & SAB(r)ab pec & 2.0 & Sy2 (Ho) & 0.71 & 1206 & 12.4 & -20.69 \\
NGC 4785 & Mu & Mz (K), DSS (R) & (R')SAB(r)ab & 3.0 & Sy2 (Veron) & 0.53 & 3735 & 49.8 & -21.70 \\
NGC 4939 & Mu & Co (K), DSS (R) & SA(s)bc & 4.0 & Sy2 (Veron) & 0.51 & 3111 & 44.3 & -22.03 \\
NGC 4941 & Mu & 2M (H), DSS (R) & (R)SAB(r)ab: & 2.0 & Sy2 (Veron) & 0.62 & 1108 & 6.4 & -17.39 \\
NGC 4968 & Mu & DSS (R) & (R')SAB0$\hat{}$0$\hat{}$ & -2.0 & Sy2 (Veron) & 0.46 & 2957 & 39.4 & -19.58 \\
NGC 5033 & Mu, Po & 2M (H), DSS (R) & SA(s)c  & 5.0 & Sy1.9 (Veron) & 0.47 & 875 & 18.7 & -21.15 \\
NGC 5135 & Mu & DSS (R) & SB(l)ab & 2.0 & Sy2 (Veron) & 0.69 & 4112 & 54.8 & -21.32 \\
NGC 5194 & Sc & ING (I), 2M (H) & SA(s)bc pec & 4.0 & Sy2 (Veron) & 0.62 & 463 & 7.7 & -20.76 \\
NGC 5273 & Po & 2M (H), DSS (R) & SA(s)0$\hat{}$0$\hat{}$ & -2.0 & Sy1.9 (Veron) & 0.91 & 1089 & 21.3 & -19.26 \\
NGC 5283 & Po & DSS (R) & S0? & -2.0 & Sy2 (Veron) & 0.91 & 2700 & 41.4 & -18.97 \\
NGC 5347 & Mu & 2M (H), DSS (R) & (R')SB(rs)ab & 2.0 & Sy2 (Veron) & 0.79 & 2335 & 36.7 & -19.72 \\
NGC 5427 & Mu & Ju (H) & SA(s)c pec & 5.0 & Sy2 (Veron) & 0.85 & 2618 & 38.1 & -21.17 \\
NGC 5548 & Ri & CF (H), DSS (R) & (R')SA(s)0/a & 0.0 & Sy1.5 (Veron) & 0.93 & 5149 & 68.7 & -21.37 \\
NGC 5643 & Mu & DSS (R) & SAB(rs)c & 5.0 & Sy2 (Veron) & 0.87 & 1199 & 16.9 & -20.91 \\
NGC 5695 & Po & 2M (H), DSS (R) & SBb & 3.0 & Sy2 (Veron) & 0.73 & 4225 & 56.3 & -20.38 \\
NGC 5929 & Mu, Po & 2M (H), DSS (R) & Sab: pec & 2.0 & Sy2 (Huchra) & 0.93 & 2492 & 38.5 & -19.93 \\
NGC 5953 & Mu & DSS (R) & SAa: pec & 1.0 & Sy2 (Veron) & 0.83 & 1965 & 33.0 & -19.29 \\
NGC 6221 & Mu & DSS (R) & SB(s)bc pec & 5.0 & Sy1 (Heckman) & 0.69 & 1482 & 19.4 & -21.67 \\
NGC 6300 & Mu & ESO (R) & SB(rs)b & 3.0 & Sy2 (Veron) & 0.66 & 1110 & 14.3 & -20.58 \\
NGC 6814 & Mu & CF (H) & SAB(rs)bc & 4.0 & Sy1.5 (Veron) & 0.93 & 1563 & 22.8 & -20.47 \\
NGC 6890 & Mu & DSS (R) & (R')SA(r:)ab & 3.0 & Sy1.9 (Veron) & 0.79 & 2419 & 31.8 & -19.69 \\
NGC 6951 & Mu & ING (I), DSS (R) & SAB(rs)bc & 4.0 & Sy2 (Veron) & 0.83 & 1424 & 24.1 & -21.20 \\
NGC 7130 & Mu & 2M (H), DSS (R) & Sa pec & 1.0 & Sy1.9 (Veron) & 0.93 & 4842 & 64.6 & -21.17 \\
NGC 7469 & Sc & CF (H), DSS (R) & (R')SAB(rs)a & 1.0 & Sy1.5 (Veron) & 0.73 & 4892 & 65.2 & -21.43 \\
NGC 7479 & St & 2M (H), DSS (R) & SB(s)c & 5.0 & Sy1.9 (Veron) & 0.76 & 2381 & 32.4 & -21.33 \\
NGC 7496 & Mu & DSS (R) & (R':)SB(rs)bc & 3.0 & Sy2 (Veron) & 0.91 & 1649 & 20.1 & -19.68 \\
NGC 7682 & Po & CF (H), DSS (R) & SB(r)ab & 2.0 & Sy2 (Veron) & 0.92 & 5134 & 68.5 & -20.51 \\
NGC 7743 & Mu & ING (I), DSS (R) & (R)SB(s)0+ & -1.0 & Sy2 (Veron) & 0.85 & 1710 & 24.4 & -19.78 \\
UGC 1395 & Po & DSS (R) & SA(rs)b & 3.0 & Sy1.9 (Veron) & 0.77 & 5208 & 69.4 & -20.35 \\
\enddata

\tablecomments{Galaxy names (col. 1), source of {\it HST} data: Mu~=~Mulchaey;
St~=~Stiavelli; Po = Pogge; Th~=~Thompson; Ri~=~Rieke; Sc~=~Scoville (col. 2), 
source of outer galaxy data: 2M = 2MASS Sky Survey; DSS~=~Digital Sky Survey;
ING~=~ING data archive; ESO~=~ESO  data archive; CF = Peletier et al. 1999;
Frei~=~Frei 1999; Mz~=~M\'arquez et al. 1999; Co~=~Combes et al. (in
preparation); Ju~=~Jungwiert et al. 1997; and band (Col 3), morphological
type (from NED; col. 4), numerical morphology ``T'' type (from RC3; col. 5),
Seyfert type and reference: Veron~=~Veron-Cetty \&  Veron 1991;
Veron82~=~Veron, Veron, \& Zuiderwijk 1981; Heckman ~=~T. Heckman, private 
communication); Ho = Ho, Filippenko, \& Sargent
1997b; GR~=~Goodrich \& Osterbrock 1983; Huchra = Huchra, Wyatt, \& Davis
1982 \& (col. 6), axial ratio (minor axis/major axis) from NED (col. 7),
heliocentric velocity in km~s$^{-1}$ (from NED; col. 8), distance in Mpc (col.
9),  absolute B-magnitude, calculated from B$_{T}$0, which was taken from RC3,
and the distance in column 9 (col. 10).}

\end{deluxetable}

\begin{deluxetable}{lcccccccc}
\tabletypesize{\scriptsize}
\tablecaption{Properties of non-Seyfert control galaxy sample \label{tab5}}
\tablewidth{0pt}
\tablehead{
\colhead{Galaxy} & \colhead{{\it HST} source} & \colhead{Other source} & \colhead{Class.} & 
\colhead{T} & \colhead{{\it b/a}} & \colhead{V$_{hel}$} & 
\colhead{D} & \colhead{M$_{B}$}    
}
\startdata
IC5267  & Mu & DSS (R) & (R)SA(rs)0/a &  0.0 & 0.74 & 1713 & 21.0 & -20.32 \\
NGC214  & Mu & DSS (R) & SAB(r)c &  5.0 & 0.74 & 4534 & 60.5 & -21.28 \\
NGC289  & St & 2M (H), DSS (R) & SAB(rs)bc &  4.0 & 0.71 & 1628 & 19.4 & -20.04 \\
NGC357  & Mu & 2M (H), DSS (R) & SB(r)0/a: &  0.0 & 0.72 & 2406 & 32.1 & -19.93 \\
NGC404  & Mu & 2M (H), DSS (R) & SA(s)0-: & -3.0 & 1.00 &  -48 &  2.4 & -15.98 \\
NGC488  & St & ING (I), DSS (R) & SA(r)b &  3.0 & 0.74 & 2272 & 29.3 & -21.43 \\
NGC628  & Mu & DSS (R) & SA(s)c &  5.0 & 0.90 &  657 &  9.7 & -20.17 \\
NGC772  & St & 2M (H), DSS (R) & SA(s)b &  3.0 & 0.59 & 2472 & 32.6 & -22.02 \\
NGC864  & Mu & ING (R), DSS (R) & SAB(rs)c &  5.0 & 0.76 & 1562 & 20.0 & -20.25 \\
NGC1345 & St & DSS (R) & SB(s)c pec: &  4.5 & 0.74 & 1529 & 18.1 & -17.49 \\
NGC1398 & Mu, St & 2M (H), DSS (R) & (R{\_}1R'{\_}2)SB(rs)ab &  2.0 & 0.76 & 1407 & 16.1 & -20.64 \\
NGC1530 & Mu & 2M (H), DSS (R) & SB(rs)b &  3.0 & 0.52 & 2461 & 36.6 & -21.40 \\
NGC1638 & Mu & 2M (H), DSS (R) & SAB(rs)0$\hat{}$0$\hat{}$?& -2.3 & 0.75 & 3320 & 44.3 & -20.46 \\
NGC1961 & Mu & ING (R), DSS (R) & SAB(rs)c &  5.0 & 0.65 & 3934 & 52.5 & -22.59 \\
NGC2179 & Mu & 2M (H), DSS (R) & (R)SA(r)0$\hat{}$+? &  0.0 & 0.69 & 2885 & 34.2 & -19.84 \\
NGC2196 & St & 2M (H), DSS (R) & (R':)SA(rs)ab &  1.0 & 0.78 & 2321 & 28.8 & -20.92 \\
NGC2223 & Mu & 2M (H), DSS (R) & SB(rs)bc &  3.0 & 0.85 & 2722 & 33.7 & -20.78 \\
NGC2276 & Mu & ING (I), DSS (R) & SAB(rs)c &  5.0 & 0.95 & 2410 & 36.8 & -21.08 \\
NGC2339 & St & DSS (R) & SAB(rs)bc &  4.0 & 0.76 & 2206 & 30.9 & -20.97 \\
NGC2344 & St & DSS (R) & SA(rs)c: &  4.5 & 0.98 &  974 & 16.0 & -18.54 \\
NGC2460 & St & ING (I), DSS (R) & SA(s)a &  1.0 & 0.76 & 1442 & 23.6 & -19.57 \\
NGC2566 & St & 2M (H), DSS (R) & (R')SB(r)ab &  2.5 & 0.68 & 1637 & 21.1 & -21.52 \\
NGC3032 & Mu & DSS (R) & SAB(r)0$\hat{}$0$\hat{}$ & -2.0 & 0.89 & 1533 & 24.5 & -19.11 \\
NGC3169 & St & ING (I), DSS (R) & SA(s)a pec &  1.0 & 0.63 & 1233 & 19.7 & -20.51 \\
NGC3277 & St & 2M (H), DSS (R)  & SA(r)ab &  2.0 & 0.89 & 1408 & 25.0 & -19.62 \\
NGC3300 & Mu & 2M (H), DSS (R) & SAB(r)0$\hat{}$0$\hat{}$:? & -2.0 & 0.53 & 3045 & 42.9 & -20.19 \\
NGC3368 & Mu & ING (I), DSS (R) & SAB(rs)ab &  2.0 & 0.69 &  897 &  8.1 & -19.74 \\
NGC3865 & Mu & 2M (H), DSS (R)  & SAB(rs)b pec: &  3.0 & 0.75 & 5702 & 76.0 & -21.84 \\
NGC3928 & St & DSS (R) & SA(s)b? &  3.0 & 1.00 &  988 & 17.0 & -18.07 \\
NGC4030 & Mu & Frei (R), 2M (H), DSS (R) & SA(s)bc &  4.0 & 0.72 & 1460 & 25.9 & -20.90 \\
NGC4143 & Mu & 2M (H), DSS (R) & SAB(s)0$\hat{}$0$\hat{}$ & -2.0 & 0.61 &  985 & 17.0 & -19.25 \\
NGC4254 & Mu & 2M (H), DSS (R) & SA(s)c &  5.0 & 0.87 & 2406 & 16.8 & -21.03 \\
NGC4260 & Mu & DSS (R) & SB(s)a &  1.0 & 0.50 & 1958 & 35.1 & -20.42 \\
NGC4384 & St & DSS (R) & Sa &  1.0 & 0.77 & 2513 & 36.6 & -19.44 \\
NGC4569 & St & 2M (H), DSS (R) & SAB(rs)ab &  2.0 & 0.46 & -235 & 16.8 & -21.34 \\
NGC4750 & St & 2M (H), DSS (R) & (R)SA(rs)ab &  2.0 & 0.91 & 1623 & 26.1 & -20.12 \\
NGC5054 & Mu & 2M (H), DSS (R) & SA(s)bc & 4.0 & 0.59 & 1741 & 27.3 & -21.05 \\
NGC5064 & Mu & DSS (R) & (R':)SA(s)ab &  2.5 & 0.46 & 2980 & 39.5 & -21.31 \\
NGC5326 & Pe & 2M (H), DSS (R) & SAa: &  1.0 & 0.50 & 2520 & 37.8 & -20.31 \\
NGC5377 & St & 2M (H), DSS (R) & (R)SB(s)a &  1.0 & 0.56 & 1793 & 31.0 & -20.52 \\
NGC5383 & Mu & Sh (K), 2M (H), DSS (R) & (R')SB(rs)b:pec &  3.0 & 0.85 & 2550 & 37.8 & -20.94 \\
NGC5448 & St & ING (I), DSS (R) & (R)SAB(r)a &  1.0 & 0.45 & 2028 & 32.6 & -20.85 \\
NGC5614 & Mu & DSS (R) & SA(r)ab pec &  2.0 & 0.80 & 3892 & 51.9 & -21.21 \\
NGC5678 & St & 2M (H), DSS (R) & SAB(rs)b &  3.0 & 0.49 & 1922 & 35.6 & -21.09 \\
NGC5739 & Mu & DSS (R) & SAB(r)0+: & -0.5 & 0.91 & 5377 & 71.7 & -21.32 \\
NGC5970 & Mu & ING (R), DSS (R) & SB(r)c &  5.0 & 0.68 & 1957 & 31.6 & -20.66 \\
NGC5985 & St & DSS (R) & SAB(r)b &  3.0 & 0.54 & 2517 & 39.2 & -21.59 \\
NGC6217 & St & 2M (H), DSS (R) & (R)SB(rs)bc &  4.0 & 0.83 & 1362 & 23.9 & -20.23 \\
NGC6340 & St & DSS (R) & SA(s)0/a &  0.0 & 0.91 & 1198 & 22.0 & -20.04 \\
NGC7096 & Mu & DSS (R) & (R')SA(rs)ab &  1.0 & 0.84 & 3100 & 36.7 & -20.18 \\
NGC7217 & St & 2M (H), DSS (R) & (R)SA(r)ab &  2.0 & 0.83 &  952 & 16.0 & -20.49 \\
NGC7280 & St & DSS (R) & SAB(r)0+ & -1.0 & 0.69 & 1844 & 26.2 & -19.25 \\
NGC7392 & Mu & 2M (H), DSS (R) & (R':)SB(rs)ab &  4.0 & 0.62 & 3192 & 42.6 & -20.90 \\
NGC7421 & St & DSS (R) & SB(r)bc &  4.0 & 0.89 & 1832 & 24.3 & -19.48 \\
NGC7716 & Mu & ING (I), DSS (R) & SAB(r)b: &  3.0 & 0.83 & 2571 & 33.7 & -20.13 \\
NGC7742 & St & ING (I), DSS (R) & SA(r)b &  3.0 & 1.00 & 1663 & 22.2 & -19.46 \\

\enddata

\tablecomments{Galaxy names (col. 1), source of {\it HST} data: Mu~=~Mulchaey;
St~=~Stiavelli; Pe = Peletier (col. 2),  source of outer galaxy data: 2M =
2MASS Sky Survey; DSS~=~Digital Sky Survey; ING~=~ING data archive; Frei~=~Frei
1999; Sh~=~Sheth et al. 2000 and band (Col 3), morphological type (from
NED; col. 4), numerical morphology ``T'' type (from RC3; col. 5), axial ratio
(minor axis/major axis) from NED (col. 6), heliocentric velocity in km~s$^{-1}$
(from NED; col. 7), distance in Mpc (col.
8),  absolute B-magnitude, calculated from B$_{T}$0, which was taken from RC3,
and the distance in column 9 (col. 9).}

\end{deluxetable}

\section{BAR PROFILES AND PROPERTIES}

The details of bar detection were described in Section 3.4. Here we present the
ellipse fits covering the whole galaxies in Figures B1 (Seyferts) and B2
(non-Seyferts), showing the ellipticities and position angles of the fitted
ellipses as a function of radius. The lengths and ellipticities of the detected
bars are tabulated in Tables~B1 (Seyferts) and B2 (non-Seyferts), together with
1~kpc in arcseconds and the galaxy diameter $D_{25}$ in arcseconds.

\setcounter{figure}{0}
\begin{figure*}[hp]
\includegraphics[width=6.0in]{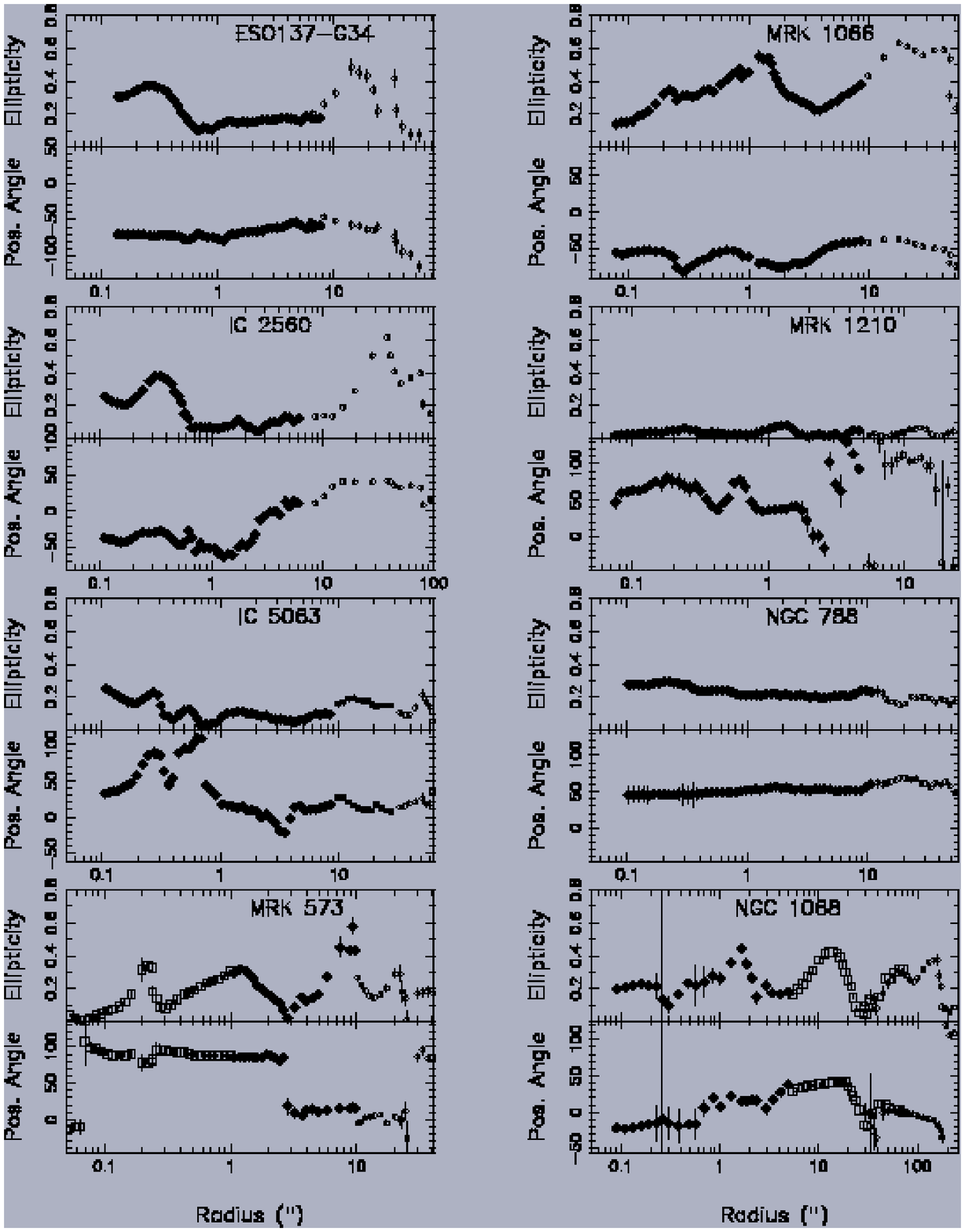}
\caption{The fitted ellipticity and position angle profiles after deprojection
for all 56 Seyfert galaxies in our sample.  The position angle here does not
necessarily have its zero point in the north direction because after
deprojection such directional distinctions may not be valid anymore. The
uncertainty bars are also shown but are often so small that they cannot be
distinguished. The {\it HST} data from NICMOS camera 2 (0.075 arcsec pixels)
are shown with  filled diamonds, NICMOS camera 1 (0.043 arcsec pixels) data are
shown with big squares, 2MASS data with big open squares, DSS data with small
open circles, and the rest of the ground-based data are shown with other
symbols.\label{fig10}} 
\end{figure*}

\setcounter{figure}{0}
\begin{figure*}[hp]
\includegraphics[width=6.0in]{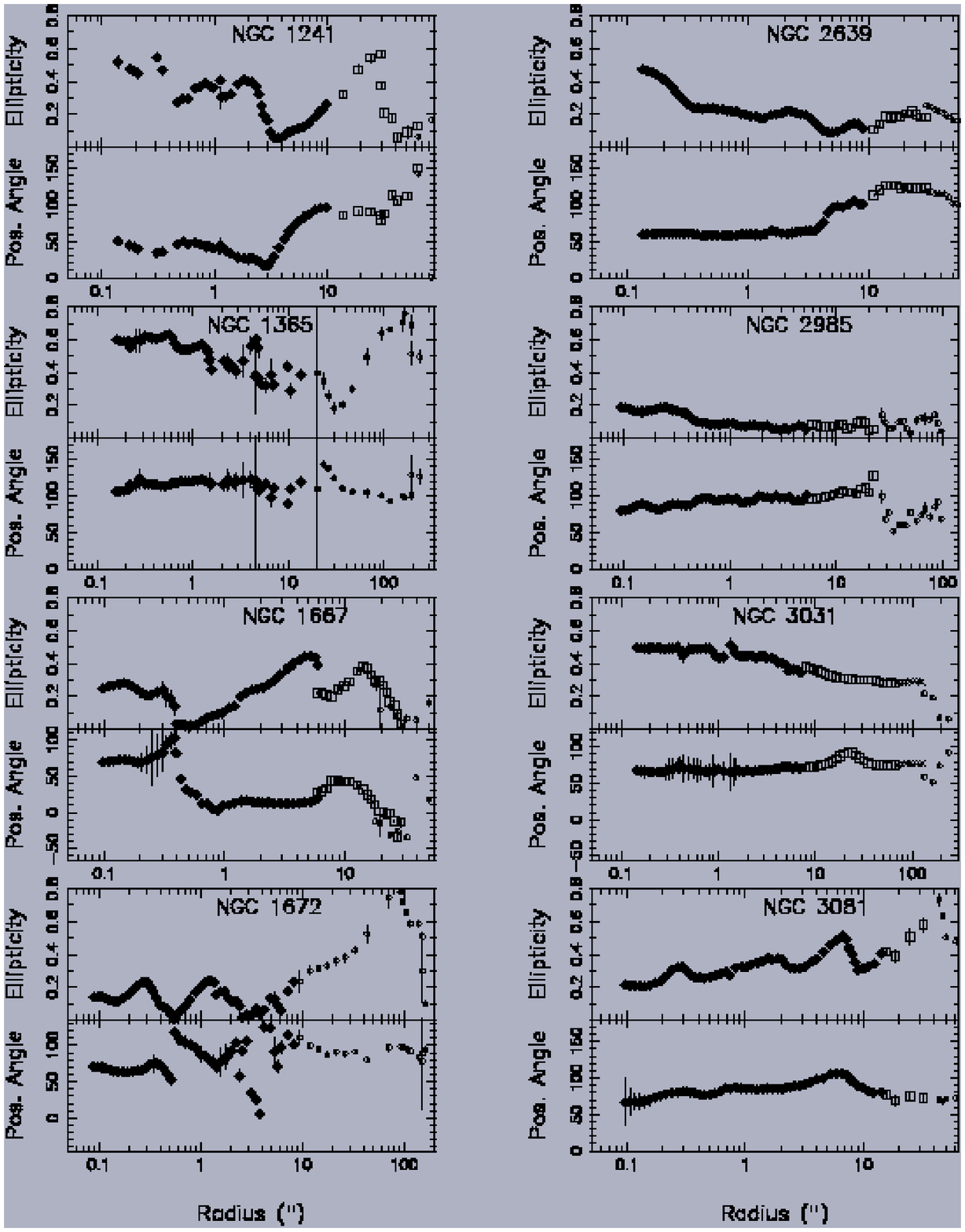}
\caption{Continued}
\end{figure*}

\setcounter{figure}{0}
\begin{figure*}[hp]
\includegraphics[width=6.0in]{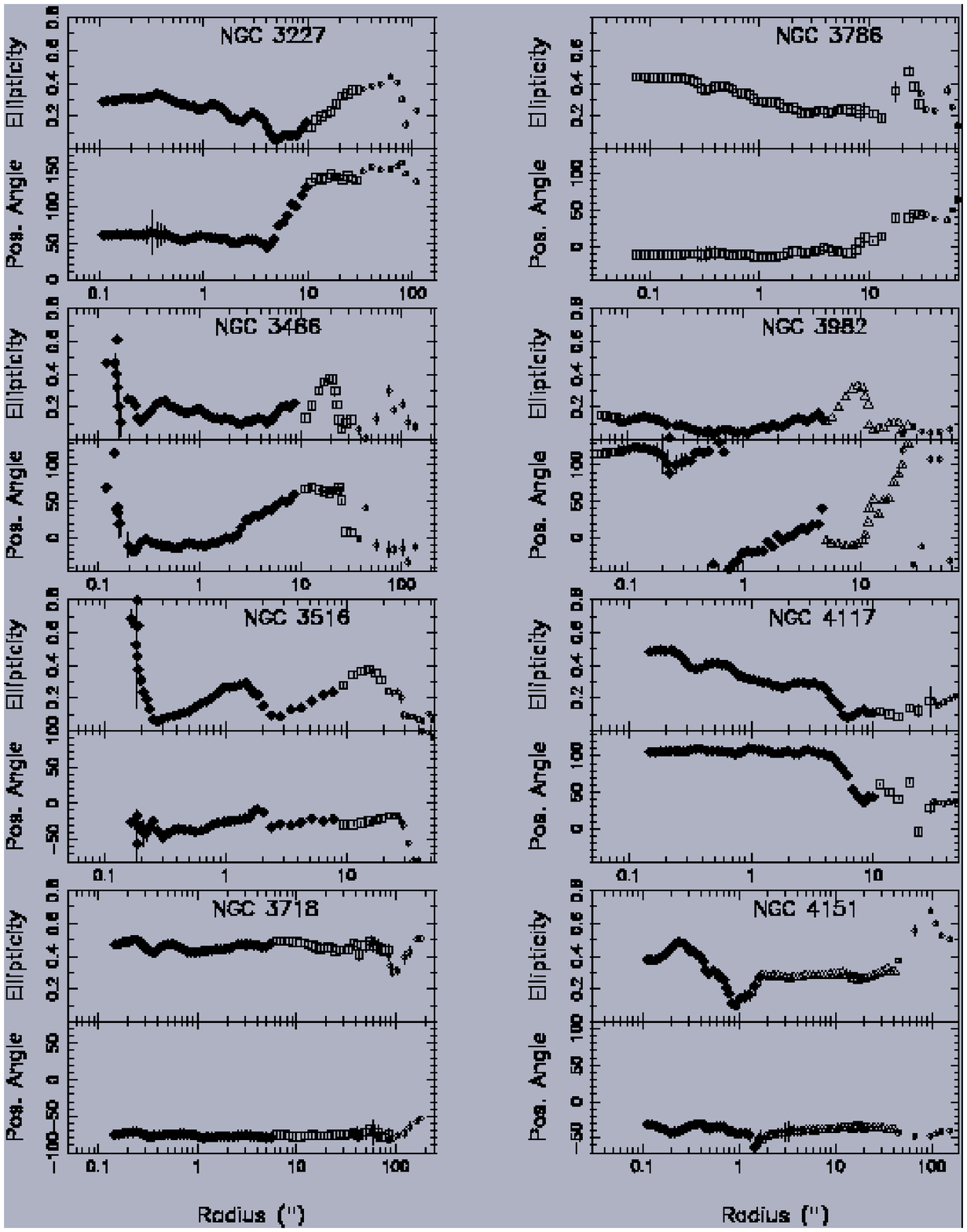}
\caption{Continued}
\end{figure*}

\setcounter{figure}{0}
\begin{figure*}[hp]
\includegraphics[width=6.0in]{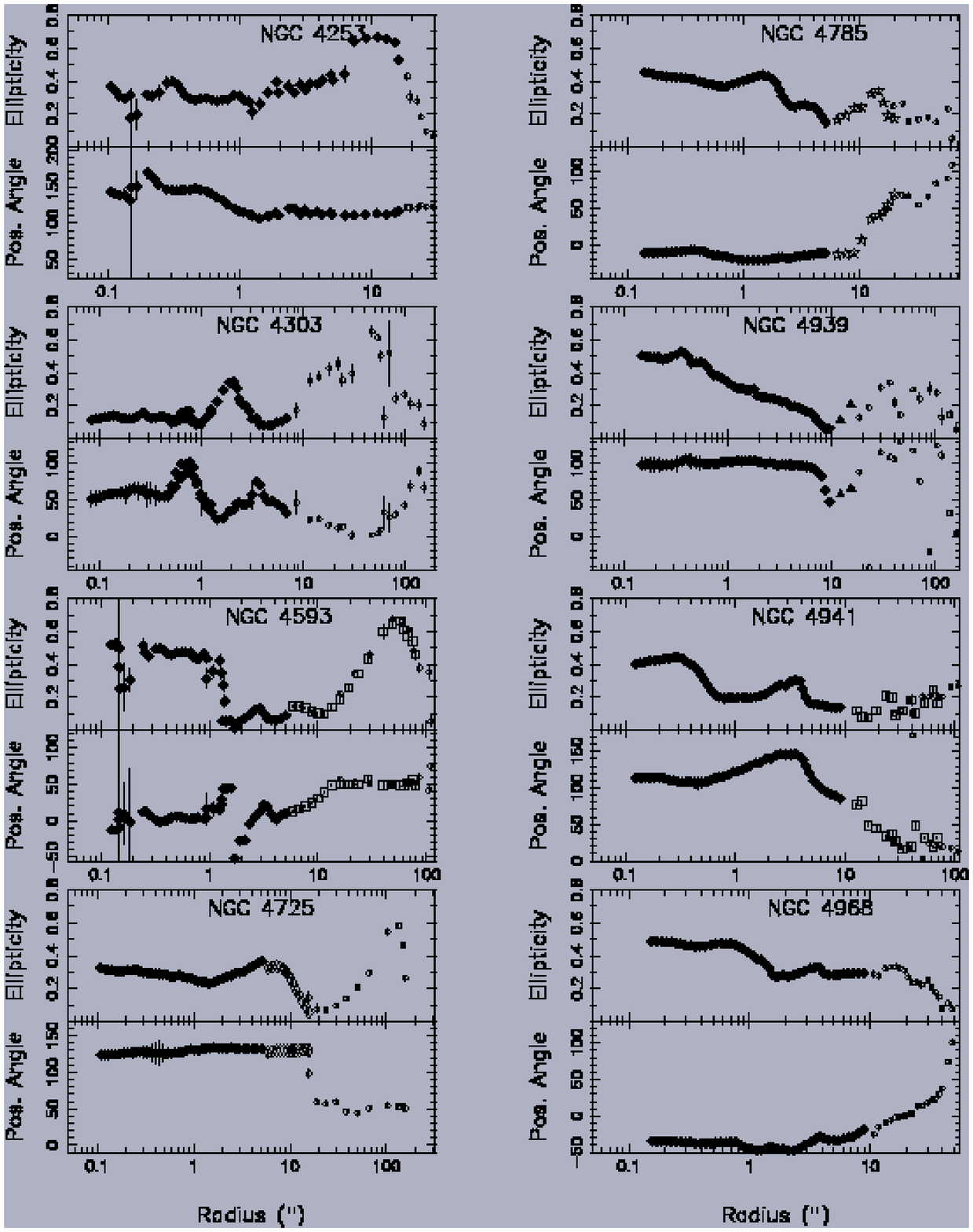}
\caption{Continued}
\end{figure*}

\setcounter{figure}{0}
\begin{figure*}[hp]
\includegraphics[width=6.0in]{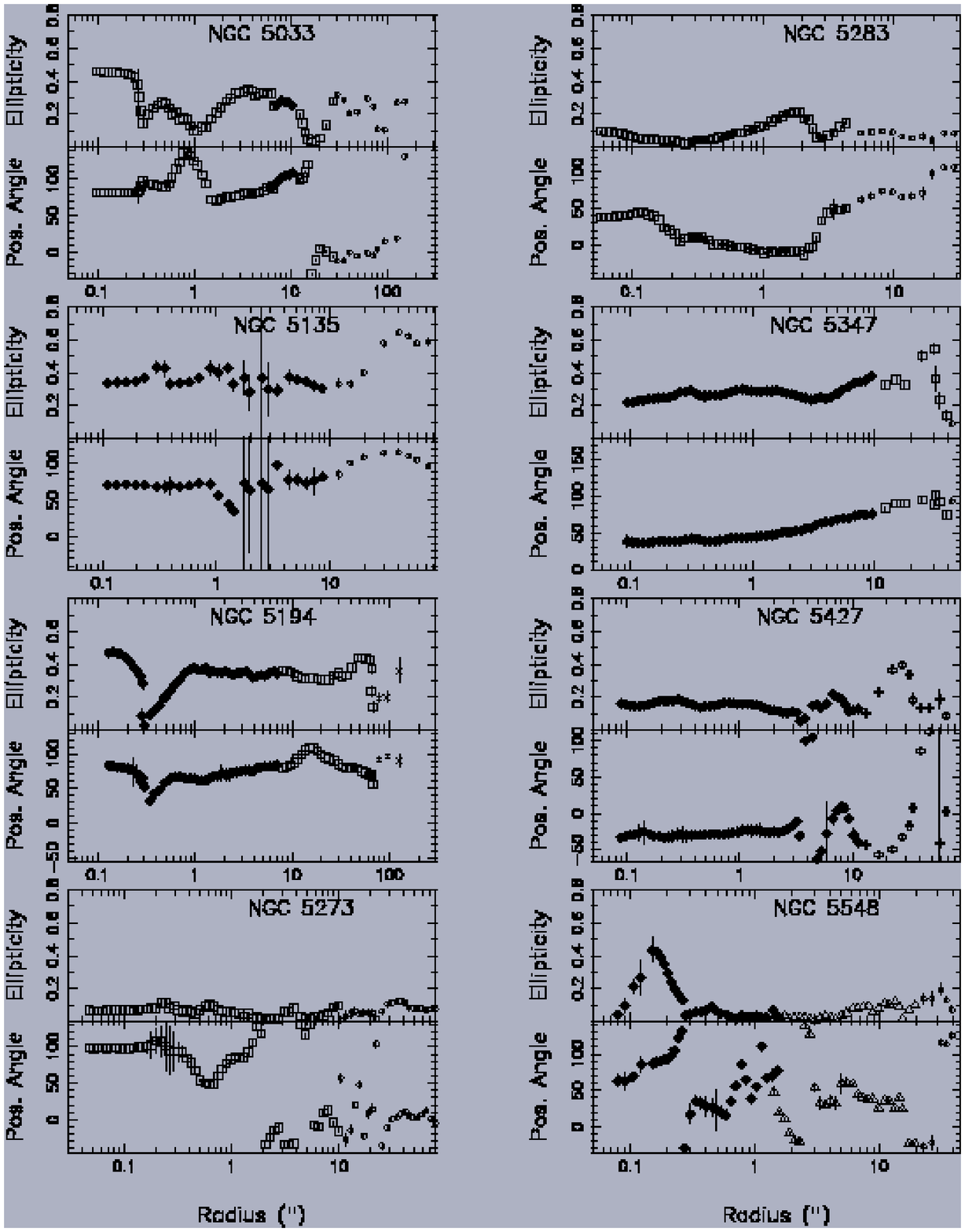}
\caption{Continued}
\end{figure*}

\setcounter{figure}{0}
\begin{figure*}[hp]
\includegraphics[width=6.0in]{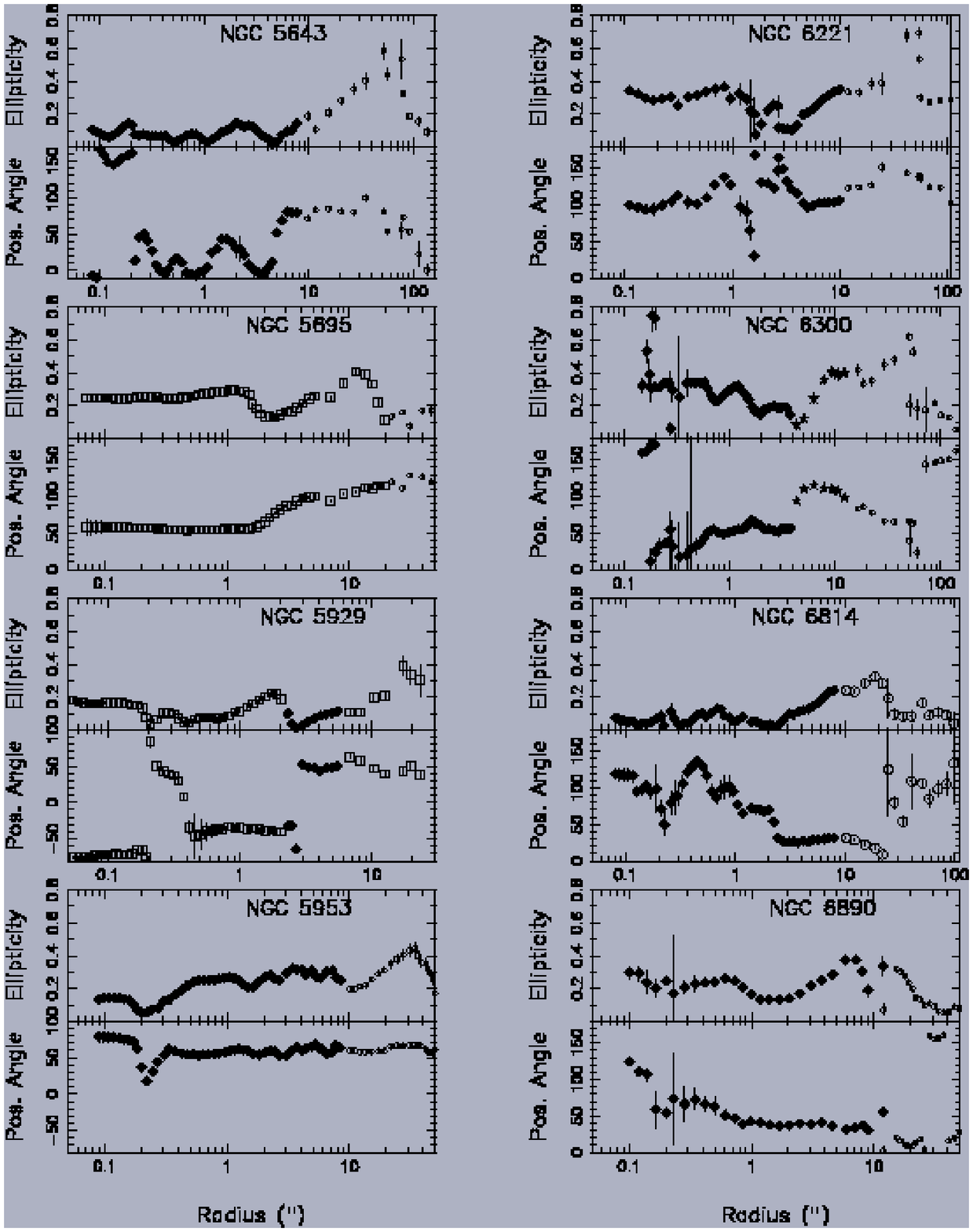}
\caption{Continued}
\end{figure*}

\setcounter{figure}{0}
\begin{figure*}[hp]
\includegraphics[width=6.0in]{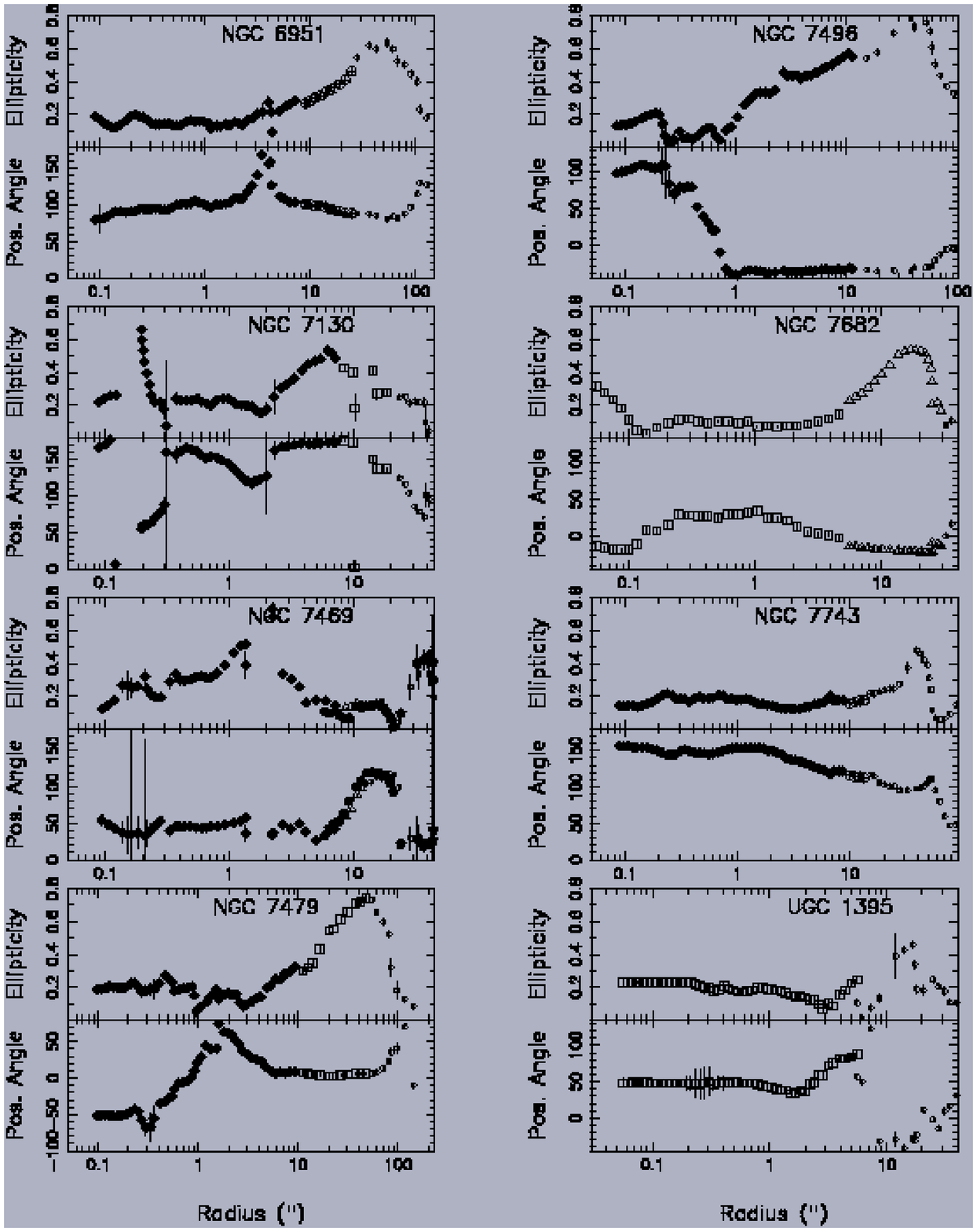}
\caption{Continued}
\end{figure*}

\begin{figure*}[hp]
\includegraphics[width=6.0in]{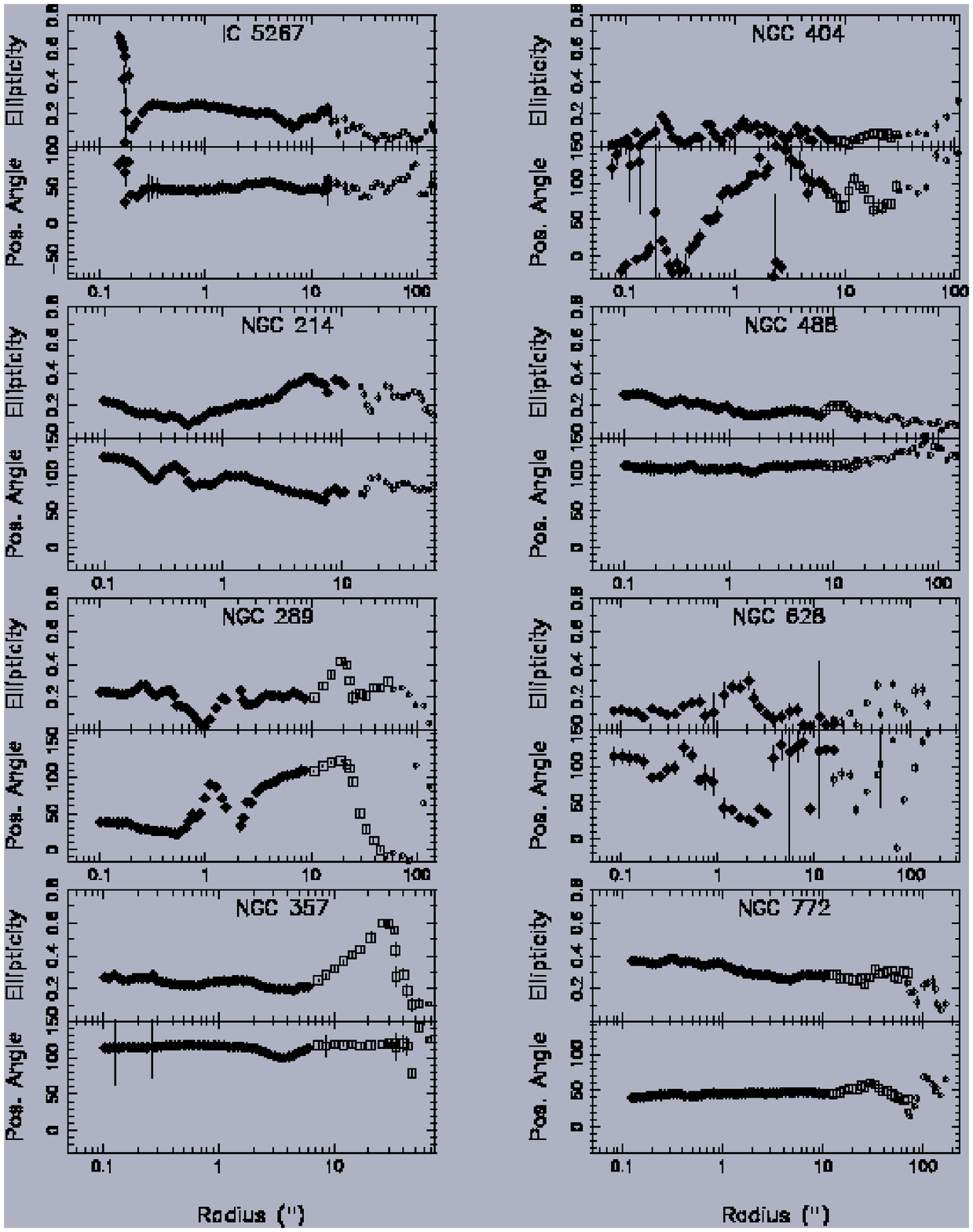}
\caption{As Figure B1, now 
for all 56 non-Seyfert galaxies in our sample. \label{fig11}}
\end{figure*}

\setcounter{figure}{1}
\begin{figure*}[hp]
\includegraphics[width=6.0in]{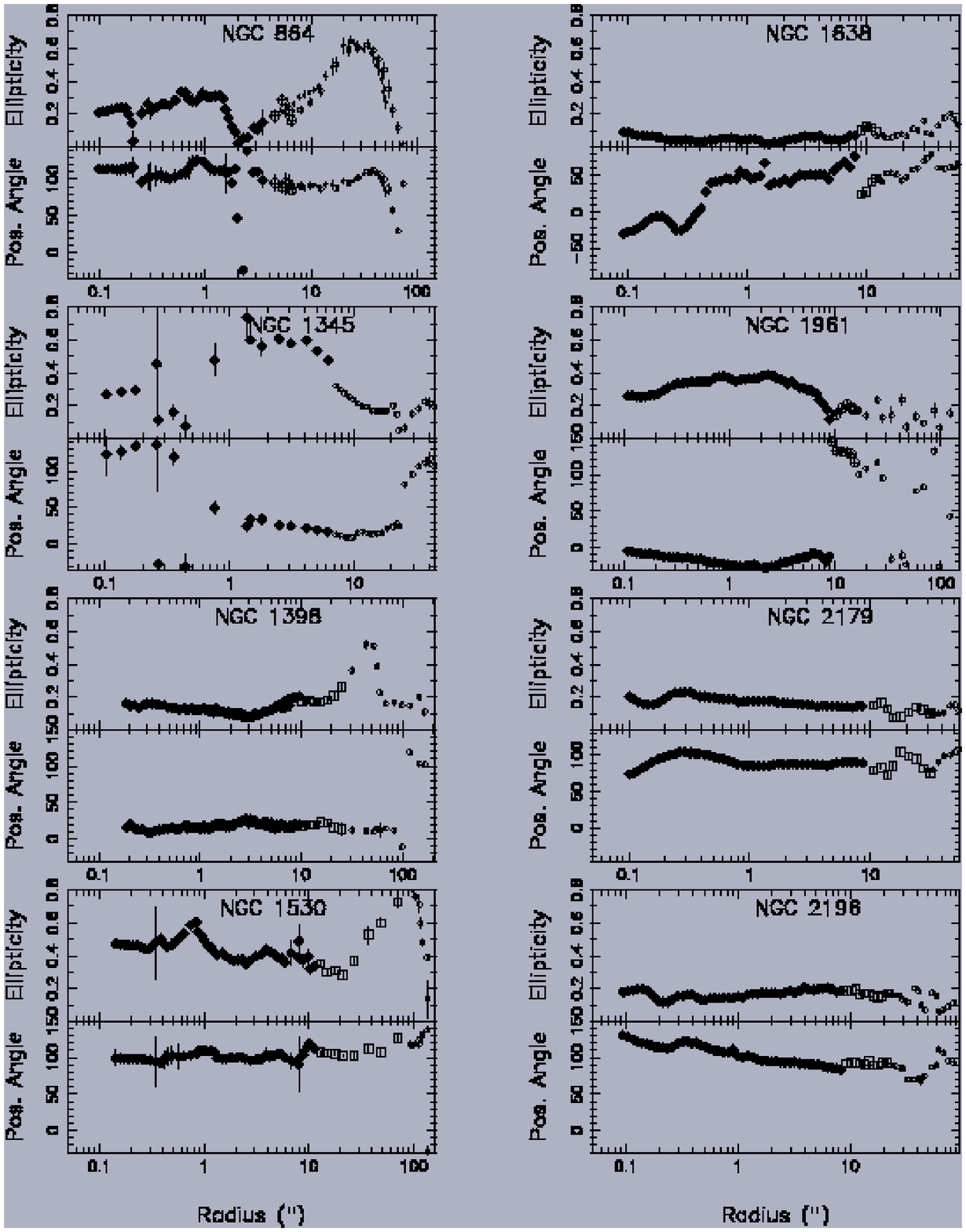}
\caption{Continued}
\end{figure*}

\setcounter{figure}{1}
\begin{figure*}[hp]
\includegraphics[width=6.0in]{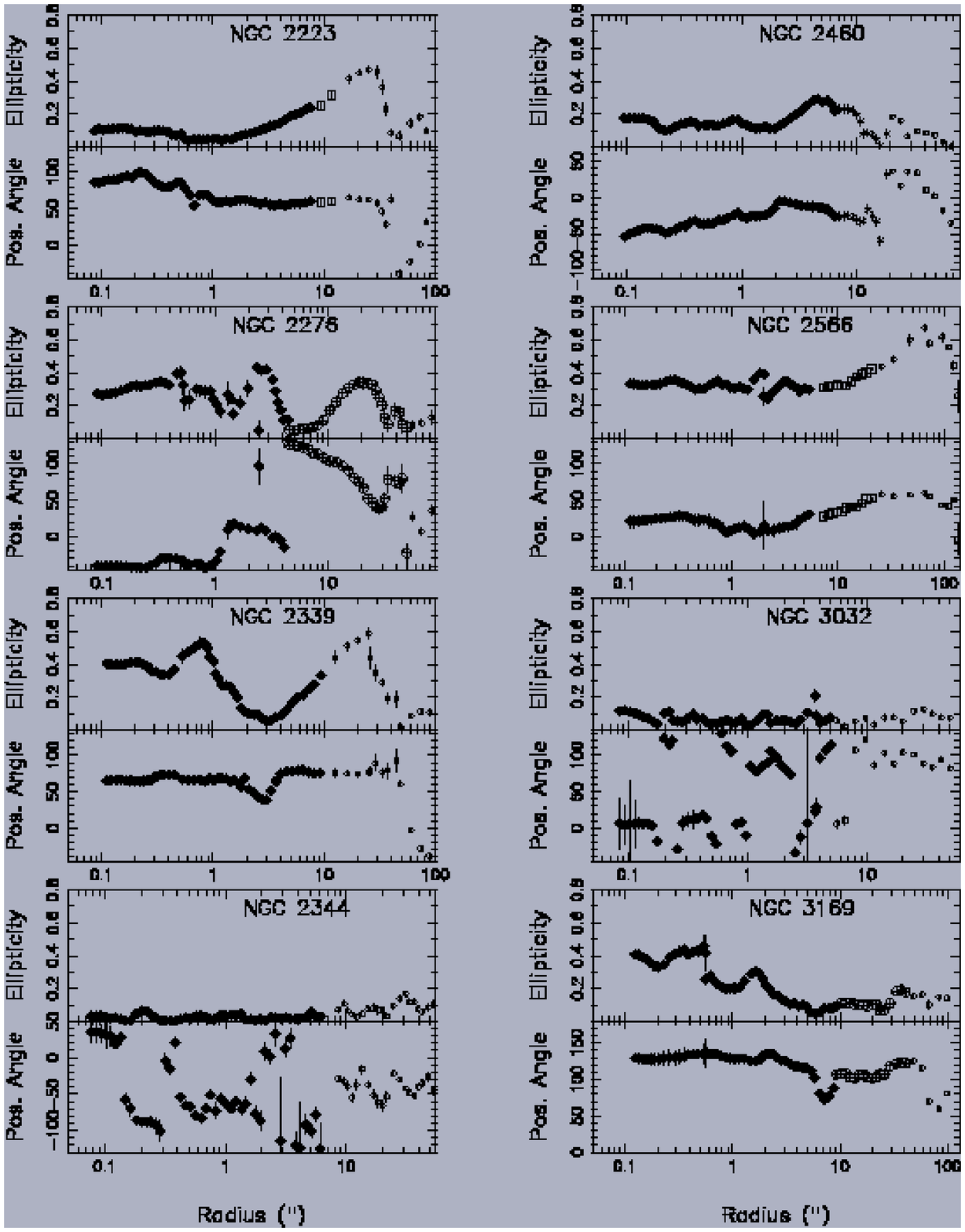}
\caption{Continued}
\end{figure*}

\setcounter{figure}{1}
\begin{figure*}[hp]
\includegraphics[width=6.0in]{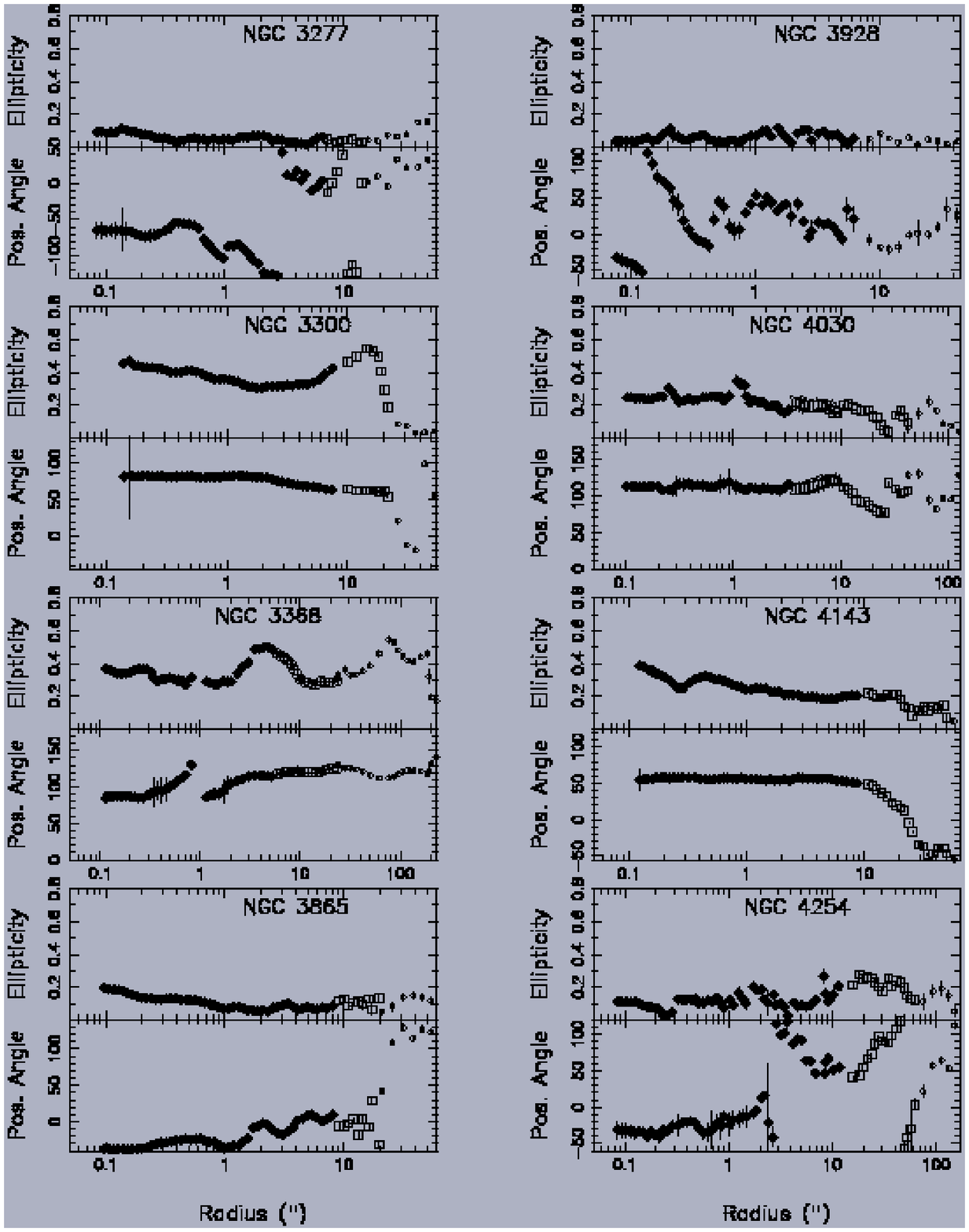}
\caption{Continued}
\end{figure*}

\setcounter{figure}{1}
\begin{figure*}[hp]
\includegraphics[width=6.0in]{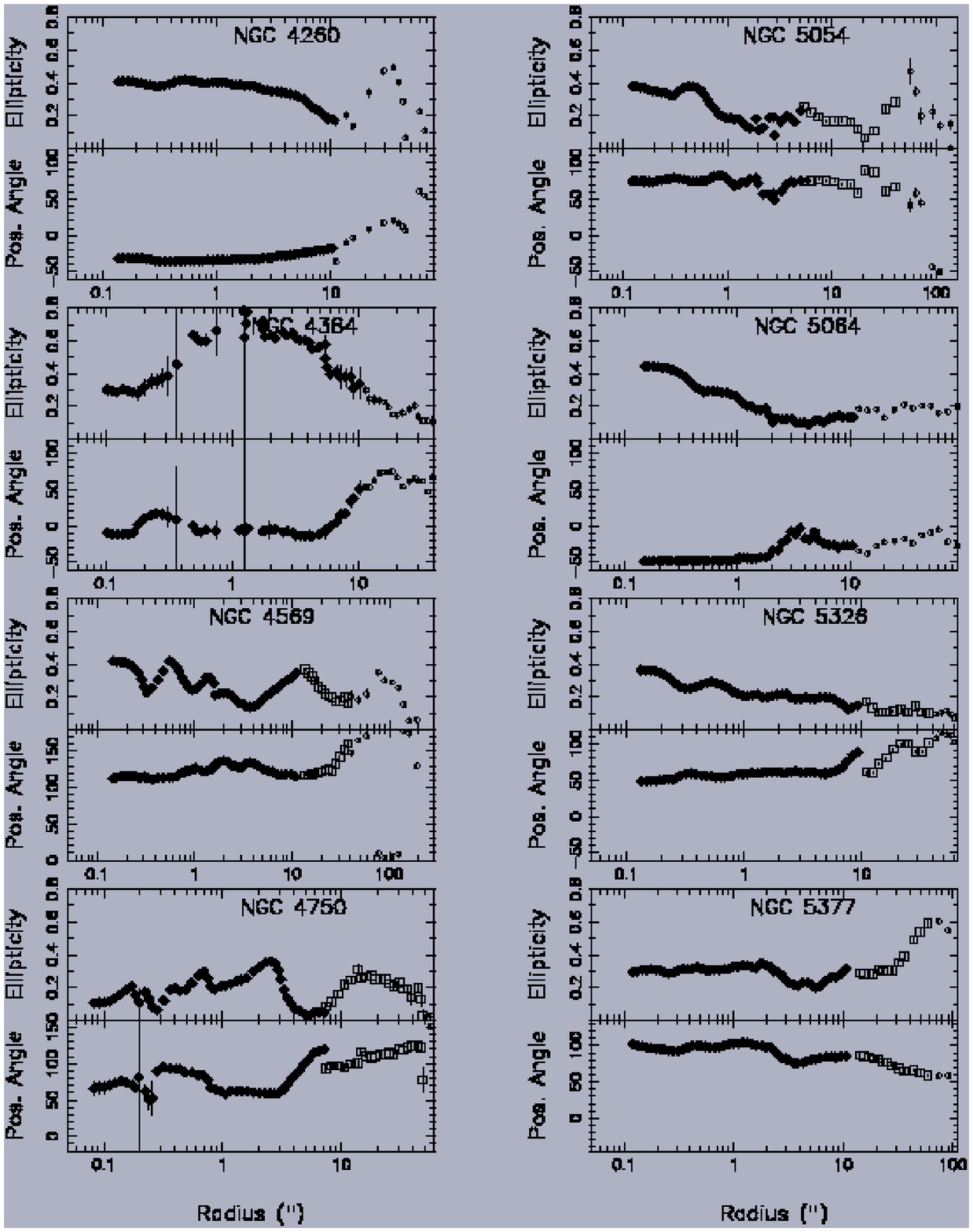}
\caption{Continued}
\end{figure*}

\setcounter{figure}{1}
\begin{figure*}[hp]
\includegraphics[width=6.0in]{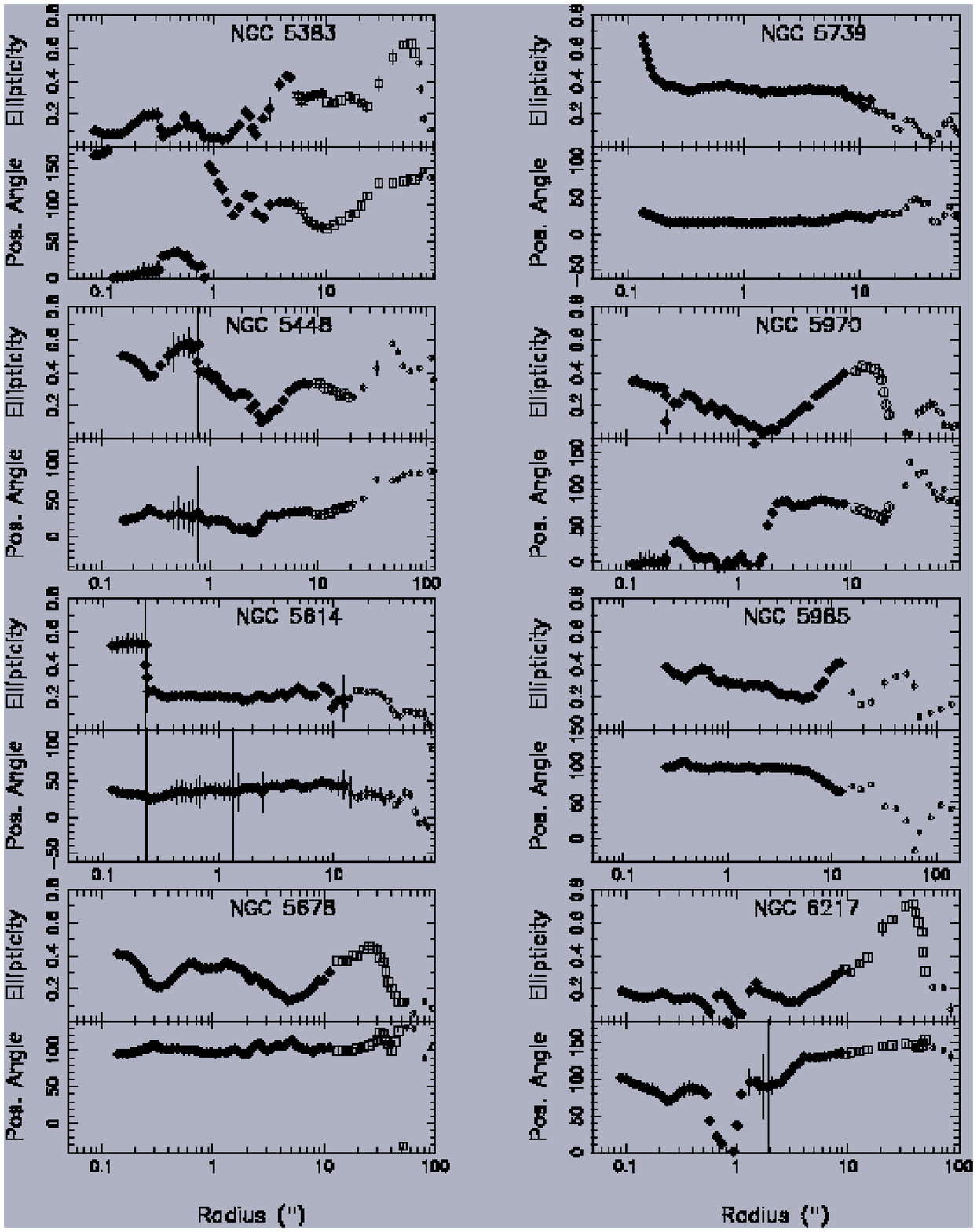}
\caption{Continued}
\end{figure*}

\setcounter{figure}{1}
\begin{figure*}[hp]
\includegraphics[width=6.0in]{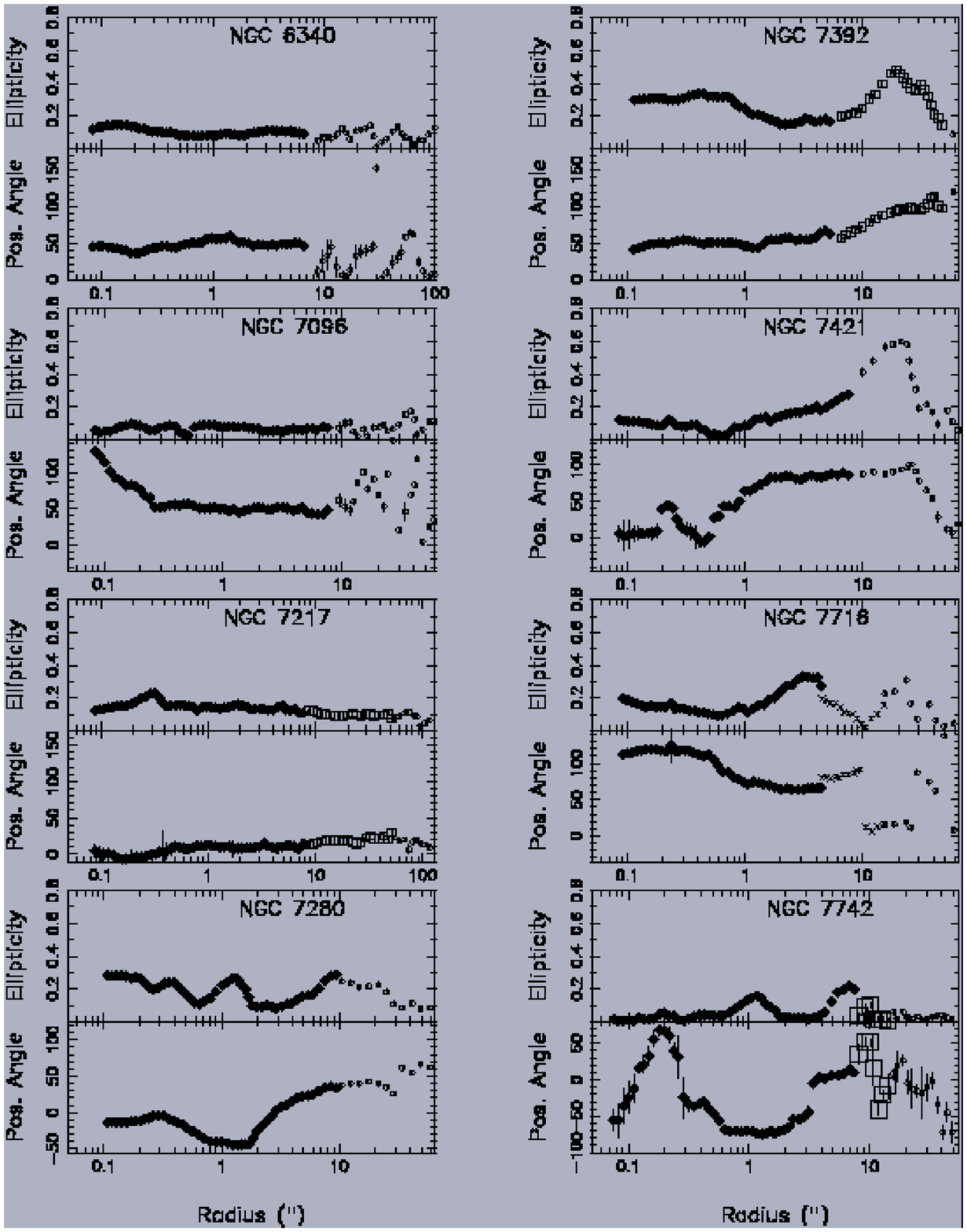}
\caption{Continued}
\end{figure*}

\clearpage

\setcounter{table}{0}
\begin{deluxetable}{lccccc}
\tablecolumns{6}
\tabletypesize{\scriptsize}
\tablecaption{Bar parameters of Seyfert galaxy sample. \label{tab6}}
\tablewidth{0pt}
\tablehead{
\colhead{Galaxy} & \colhead{1 kpc in arcsec} & \multicolumn{2}{c}{Deprojected bar radii} 
 & \colhead{Deprojected bar} & \colhead{Galaxy diameter} \\
\colhead{} & \colhead{} & \colhead{arcsec} & \colhead{pc} & \colhead{ellipticities} &
\colhead{arcsec}}
\startdata 
ESO 137-G34 & 5.6 & 14 & 2500 & 0.48 & 154 \\
IC 2560 &  5.3  & 38 & 7200 & 0.61 & 203 \\
IC 5063 &  4.5  & & & & 128 \\
Mrk 573 &  3.0  & 1.2, 9.3, 22 & 400, 3100, 7300 & 0.32, 0.58, 0.29 & 83 \\
Mrk 1066 & 4.3  & 1.2, 17 & 280, 4000 & 0.55, 0.63 & 109 \\
Mrk 1210 & 3.8  & & & & 50 \\
NGC 788 &  3.8  & & & & 114 \\
NGC 1068 & 14.3 & 1.7, 14 & 120, 1000 & 0.44, 0.42 & 425 \\
NGC 1241 & 7.8  & 1.8, 30 & 230, 3800 & 0.41, 0.56 & 177 \\
NGC 1365 & 12.2 & 4.7, 165 & 390, 13500 & 0.60, 0.76 & 673 \\
NGC 1667 & 3.4  & 5.2, 14 & 1400, 3800 & 0.45, 0.38 & 112 \\
NGC 1672 & 14.2 & 92 & 6500 & 0.78 & 396 \\
NGC 2639 & 4.6  & & & & 112 \\
NGC 2985 & 9.2  & & & & 274 \\
NGC 3031 & 147 & & & & 1653 \\
NGC 3081 & 6.3 & 6.7, 44 & 1100, 7000 & 0.51, 0.73 & 128 \\
NGC 3227 & 10.0 & 62 & 6200 & 0.44 & 330 \\
NGC 3486 & 27.9 & 19 & 680 & 0.38 & 425 \\
NGC 3516 & 5.3  & 1.5, 15 & 280, 2800 & 0.29, 0.38 & 104 \\
NGC 3718 & 12.1 & & & & 488 \\
NGC 3786 & 5.0  & 23 & 4600 & 0.47 & 131 \\
NGC 3982 & 12.1 & 9.5 & 790 & 0.33 & 141 \\
NGC 4117 & 12.1 & & & & 97 \\
NGC 4151 & 10.2 & 97 & 9500 & 0.68 & 379 \\
NGC 4253 & 3.6  & 11 & 3100 & 0.67 & 59 \\
NGC 4303 & 13.6 & 2.1, 47 & 150, 3500 & 0.34, 0.65 & 387 \\
NGC 4593 & 5.2  & 48 & 9200 & 0.67 & 233 \\
NGC 4725 & 16.6 & 5.1, 134 & 310, 8100 & 0.37, 0.58 & 643 \\
NGC 4785 & 4.1  & & & & 144 \\
NGC 4939 & 4.7  & & & & 337 \\
NGC 4941 & 32.2 & 3.5 & 110 & 0.30 & 218 \\
NGC 4968 & 5.2  & & & & 109 \\
NGC 5033 & 11.0 & 0.5, 3.7, 30 & 40, 340, 2700 & 0.27, 0.35, 0.32 & 643 \\
NGC 5135 & 3.8  & 40 & 10500 & 0.65 & 165 \\
NGC 5194 & 26.8 & & & & 673 \\
NGC 5273 & 9.7  & & & & 161 \\
NGC 5283 & 5.0  & 1.7 & 340 & 0.22 & 64 \\
NGC 5347 & 5.6  & 31 & 5500 & 0.55 & 102 \\
NGC 5427 & 5.4  & & & & 173 \\
NGC 5548 & 3.0  & & & & 87 \\
NGC 5643 & 12.2 & 52 & 4300 & 0.58 & 308 \\
NGC 5695 & 3.7  & 11 & 3000 & 0.41 & 93 \\
NGC 5929 & 5.4  & 1.7 & 310 & 0.23 & 59 \\
NGC 5953 & 6.3  & 34 & 5400 & 0.45 & 100 \\
NGC 6221 & 10.6 & 54 & 5100 & 0.69 & 256 \\
NGC 6300 & 14.4 & 1.2, 50 & 80, 3500 & 0.32, 0.62 & 294 \\
NGC 6814 & 9.0  & 19 & 2100 & 0.32 & 208 \\
NGC 6890 & 6.5  & 7.1 & 1100 & 0.38 & 93 \\
NGC 6951 & 8.6  & 53 & 6200 & 0.64 & 287 \\
NGC 7130 & 3.2  & 6.2 & 1900 & 0.53 & 91 \\
NGC 7469 & 3.2  & 2.2 & 690 & 0.52 & 91 \\
NGC 7479 & 6.4  & 48 & 7500 & 0.74 & 256 \\
NGC 7496 & 10.3 & 38 & 3700 & 0.79 & 199 \\
NGC 7682 & 3.0  & 17 & 5700 & 0.54 & 77 \\
NGC 7743 & 8.4  & 39 & 4600 & 0.48 & 181 \\
UGC 1395 & 3.0 & 17 & 5700 & 0.46 & 77 \\
\enddata

\tablecomments{Galaxy names (col. 1), arcsec in the galaxy image corresponding to 1 kpc (col. 2),
galaxy diameter at the 25 mag arcsec$^{-2}$ B-mag level, in arcsec, corrected
for the inclination, redshift, and Galactic absorption, from RC3 (col. 6).}

\end{deluxetable}

\clearpage 

\begin{deluxetable}{lcccccccc}
\tablecolumns{6}
\tabletypesize{\scriptsize}
\tablecaption{Bar parameters of non-Seyfert control galaxy sample \label{tab7}}    
\tablewidth{0pt}                                                        
\tablehead{
\colhead{Galaxy} & \colhead{1 kpc in arcsec} & \multicolumn{2}{c}{Deprojected bar radii} 
 & \colhead{Deprojected bar} & \colhead{Galaxy diameter} \\
\colhead{} & \colhead{} & \colhead{arcsec} & \colhead{pc} & \colhead{ellipticities} &
\colhead{arcsec} \\   
}
\startdata
IC5267  & 9.8 & & & & 301 \\
NGC214  & 3.4 & & & & 114 \\
NGC289  & 10.6 & 19 & 1800 & 0.42 & 315 \\
NGC357  & 6.4 & 27 & 4200 & 0.60 & 151 \\
NGC404  & 85.9 & & & & 223 \\
NGC488  & 7.0 & & & & 322 \\
NGC628  & 21.3 & 2.1 & 100 & 0.30 & 643 \\
NGC772  & 6.3 & & & & 455 \\
NGC864  & 10.3 & & & & 287 \\
NGC1345 & 11.4 & 1.4 & 120 & 0.73 & 93 \\
NGC1398 & 12.8 & 44 & 3400 & 0.52 & 425 \\
NGC1530 & 5.6 & 0.8, 92 & 150, 16400 & 0.61, 0.86 & 315 \\
NGC1638 & 4.7 & & & & 120 \\
NGC1961 & 3.9 & & & & 301 \\
NGC2179 & 6.0 & & & & 107 \\
NGC2196 & 7.2 & & & & 185 \\
NGC2223 & 6.1 & 25 & 4100 & 0.47 & 208 \\
NGC2276 & 5.6 & 2.4 & 430 & 0.43 & 177 \\
NGC2339 & 6.7 & 0.8, 25 & 90, 3700 & 0.53, 0.59 & 185 \\
NGC2344 & 12.9 & & & & 112 \\
NGC2460 & 8.7 & 4.6 & 530 & 0.29 & 154 \\
NGC2566 & 9.8 & 66 & 6700 & 0.67 & 281 \\
NGC3032 & 8.4 & & & & 117 \\
NGC3169 & 10.5 & 1.6 & 150 & 0.30 & 262 \\
NGC3277 & 8.3 & & & & 117 \\
NGC3300 & 4.8 & 14 & 2900 & 0.55 & 107 \\
NGC3368 & 25.5 & 4.8, 77 & 190, 3000 & 0.50, 0.54 & 455 \\
NGC3865 & 2.7 & & & & 125 \\
NGC3928 & 12.1 & & & & 91 \\
NGC4030 & 8.0 & & & & 256 \\
NGC4143 & 12.1 & & & & 128 \\
NGC4254 & 12.3 & & & & 337 \\
NGC4260 & 5.9 & 35 & 5900 & 0.49 & 161 \\
NGC4384 & 5.6 & 1.9 & 340 & 0.81 & 77 \\
NGC4569 & 12.3 & 14 & 1100 & 0.37 & 586 \\
NGC4750 & 7.9 & 2.5, 14 & 320, 1800 & 0.36, 0.31 & 123 \\
NGC5054 & 7.6 & & & & 315 \\
NGC5064 & 5.2 & & & & 173 \\
NGC5326 & 5.5 & & & & 131 \\
NGC5377 & 6.7 & 74 & 11000 & 0.61 & 223 \\
NGC5383 & 5.5 & 4.4, 58 & 800, 10500 & 0.43, 0.63 & 190 \\
NGC5448 & 6.3 & 49 & 7800 & 0.58 & 239 \\
NGC5614 & 4.0 & & & & 147 \\
NGC5678 & 5.8 & 25 & 4300 & 0.46 & 199 \\
NGC5739 & 2.9 & & & & 134 \\
NGC5970 & 6.5 & 12 & 1800 & 0.44 & 177 \\
NGC5985 & 5.3 & & & & 330 \\
NGC6217 & 8.6 & 39 & 4500 & 0.72 & 185 \\
NGC6340 & 9.4 & & & & 203 \\
NGC7096 & 5.6 & & & & 114 \\
NGC7217 & 12.9 & & & & 256 \\
NGC7280 & 7.9 & 1.3 & 160 & 0.27 & 128 \\
NGC7392 & 4.8 & & & & 131 \\
NGC7421 & 8.5 & 21 & 2500 & 0.60 & 125 \\
NGC7716 & 6.1 & 3.1, 23 & 510, 3800 & 0.33, 0.31 & 131 \\
NGC7742 & 9.3 & 1.2, 6.8 & 130, 730 & 0.16, 0.22 & 107 \\
\enddata
\tablecomments{Galaxy names (col. 1), arcsec in the galaxy image corresponding to 1 kpc (col. 2),
galaxy diameter at the 25 mag arcsec$^{-2}$ B-mag level, in arcsec, corrected
for the inclination, redshift, and absorption, from RC3 (col. 6).}

\end{deluxetable}


\clearpage

\begin{thebibliography}{}
\bibitem[Adams(1977)]{ada} Adams, T. F. 1977, \apjs, 33, 19
\bibitem[Antonucci(1993)]{ant} Antonucci, R. 1993, \araa, 31, 473
\bibitem[Athanassoula(1992)]{ath92} Athanassoula, E. 1992, \mnras, 259, 345
\bibitem[Athanassoula \& Martinet(1982)]{ath} Athanassoula, E., \& Martinet, L.
   1980, \aap, 87, L10
\bibitem[Balick \& Heckman(1982)]{bal} Balick, B., \& Heckman, T. M. 1980, 
   \araa, 20, 431
\bibitem[Buta \& Combes(1996)]{buta0} Buta, R., \& Combes, F. 1996, Fund. Cosmic
   Phys., 17, 95
\bibitem[Buta \& Crocker(1991)]{buta1} Buta, R., \& Crocker, D. 1991, \aj, 102,
   1715
\bibitem[Buta \& Crocker(1993)]{buta2} Buta, R., \& Crocker, D. 1993, \aj, 105,
   1344
\bibitem[Colina \& Wada(2000)]{col} Colina, L., \& Wada, K. 2000, \apj, 529, 845  
\bibitem[Combes \& Gerin (1985)]{com2} Combes, F., \& Gerin, M. 1985, \aap, 150,
   327
\bibitem[Dahari(1984)]{dah} Dahari, O. 1984, \aj, 89, 966
\bibitem[de Robertis et al.(1998)]{rob} de Robertis, M. M., Yee, H. K. C., \&
   Hayhoe, K. 1998, \apj, 496, 93
\bibitem[de Vaucouleurs(1974)]{vau} de Vaucouleurs, G. (1974), in IAU Symp. 58,
   The Formation and Dynamics of Galaxies, ed. J. R. Shakeshaft, (Dordrecht:
   Reidel), p. 335
\bibitem[de Vaucouleurs et al.(1991)]{vau2} de Vaucouleurs G., de Vaucouleurs A., 
   Corwin H. G. Jr., Buta R. J., Paturel G., \& Fouque, P., 1991, Third 
   Reference Catalogue of Bright Galaxies (New York: Springer)
\bibitem[Devereux et al.(1992)]{dev} Devereux, N. A., Kenney, J. D. P., \& 
   Young, J. S. 1992, \aj, 103, 784
\bibitem[Elmegreen et al.(1998)]{elm1} Elmegreen, D. M., Chromey, F. R., \& 
   Santos, M. 1998, \aj, 116, 1221
\bibitem[Elmegreen et al.(1990)]{elm2} Elmegreen, D. M., Elmegreen, B. G., \& 
   Bellin, A. D. 1990, \apj, 364, 415
\bibitem[Emsellem \& Ferruit(2000)]{ems} Emsellem, E., \& Ferruit, R. 2000,
   \aap, 357, 111 
\bibitem[Erwin \& Sparke(1999)]{erw} Erwin, P., \& Sparke, L. 1999a, in ASP
   Conf. Ser. 182, Galaxy Dynamics, ed. D. Merritt, J. A. Sellwood, \& M. 
   Valluri, (San Francisco: ASP), 243
\bibitem[Erwin \& Sparke(1999)]{erw2} Erwin, P., \& Sparke, L. 1999b, \apjl,
   521, L37
\bibitem[Eskridge et al.(2000)]{esk} Eskridge, P. B. et al. 2000, \aj, 119, 536
\bibitem[Forbes et al.(1994)]{for} Forbes, D. A., Kotilainen, J. K., \& Moorwood,
   A. F. M. 1994, \apjl, 433, L13
\bibitem[Frei (1999)]{frei} Frei, Z. 1999, \apss, 269, 649
\bibitem[Friedli et al.(1996)]{fri2} Friedli, D., Wozniak, H., Rieke, M., 
   Martinet, L., \& Bratschi, P. 1996, \aaps, 118, 461
\bibitem[Fuentes-Williams \& Stocke(1988)]{fue} Fuentes-Williams, T. \& Stocke,
   J. T. 1988, \aj, 96, 1235
\bibitem[Giovanelli et al.(1994)]{gio} Giovanelli, R., Haynes, M. P., Salzer, 
   J. J., Wegner, G., da Costa, L. N., \& Freudling, W. 1994, \aj, 107, 2036 
\bibitem[Gonz\'alez Delgado et al.(2001)]{gon} Gonz\'alez Delgado, R. M.,
   Heckman, T., \& Leitherer, C. 2001, \apj, 546, 845
\bibitem[Goodrich \& Osterbrock(1983)]{god} Goodrich, R. W., \& Osterbrock, D.
   E. 1983, \apj, 269, 416
\bibitem[Greusard et al.(2000)]{gre} Greusard, D., Friedli, D., Wozniak, H.,
   Martinet, L., \& Martin, P. 2000, \aaps, 145, 425
\bibitem[Grosb{\o}l(2001)]{gros} Grosb{\o}l, P. 2001, Talk at the INAOE workshop
   on Disk Galaxies, Puebla, Mexico
\bibitem[Ho et al.(1997a)]{ho} Ho, L. C., Filippenko, A. V., \& Sargent, W. L. W.
   1997a, \apj, 487, 591
\bibitem[Ho et al.(1997b)]{ho2} Ho, L. C., Filippenko, A. V., \& Sargent, W. L. W.
   1997b, \apjs, 112, 315
\bibitem[Huchra \& Burg(1992)]{huc} Huchra, J., \& Burg, R. 1992, \apj, 393, 90
\bibitem[Huchra et al.(1982)]{huc2} Huchra, J., Wyatt, W. F., \& Davis, M. 1982, 
   \aj, 87, 1682
\bibitem[Ishizuki et al.(1990)]{ish} Ishizuki, S., Kawabe, R., Ishiguro, M.
   Okumura, S. K., \& Morita, K.--I. 1990, \nat, 344, 224
\bibitem[Jogee(1998)]{jog2} Jogee, S., Kenney, J. D. P., \& Smith, B. J. 1998,
   \apjl, 494, L185
\bibitem[Jogee(1999)]{jog3} Jogee, S., Kenney, J. D. P., \& Smith, B. J. 1999,
   \apj, 526, 665
\bibitem[J{\o}rgensen et al.(1991)]{jor} J{\o}rgensen, I., Franx, M., \&
   Kjaergaard, P. 1992, \aaps, 95, 489
\bibitem[Jungwiert et al.(1997)]{jung} Jungwiert, B., Combes, F., \& Axon, D. J.
   1997, \aaps, 125, 479 
\bibitem[Knapen et al.(1995a)]{kna1} Knapen, J. H.,  Beckman, J. E., Shlosman, 
   I., Peletier, R. F., Heller, C. H., \& de Jong, R. S. 1995a, \apjl, 443,
   L73   
\bibitem[Knapen et al.(1995b)]{kna2} Knapen, J. H., Beckman, J. E., Heller,
   C. H., Shlosman, I., \& de Jong, R. S. 1995a, \apj, 454,
   623   
\bibitem[Knapen et al.(2000)]{kna3} Knapen, J. H., Shlosman, I., \&
    Peletier, R. F. 2000, \apj, 529, 93 (KSP)
\bibitem[Kormendy(1982)]{kor} Kormendy, J. 1982, \apj, 257, 75 
\bibitem[Kotilainen et al.(2000)]{kot} Kotilainen, J. K., Reunanen, J.,
   Laine, S., \& Ryder, S. D. 2000, \aap, 353, 834
\bibitem[Maciejewski \& Sparke(2000)]{mac} Maciejewski, W., \& Sparke, L. 2000,
   \mnras, 313, 745
\bibitem[MacKenty(1989)]{mack} MacKenty, J. W. 1989, \apj, 343, 125
\bibitem[Maiolino et al.(2000)]{mai} Maiolino, R., Alonso-Herrero, A.,
   Anders, S., Quillen, A., Rieke, M. J., Rieke, G. H., \& Tacconi-Garman, L. E.
   2000, \apj, 531, 219
\bibitem[M\'arquez et al.(1999)]{marq} M\'arquez, I. et al. 1999, \aaps, 
   140, 1
\bibitem[M\'arquez et al.(2000)]{marq2} M\'arquez, I. et al. 2000, \aap, 360, 431
\bibitem[Martin(1995)]{mart} Martin, P. 1995, \aj, 109, 2428
\bibitem[Martinet \& Friedli(1997)]{mfri} Martinet, L., \& Friedli, D. 1997, 
   \aap, 323, 363
\bibitem[Martini \& Pogge(1999)]{mar} Martini, P., \& Pogge, R. W. 1999, \aj,
   118, 2646
\bibitem[Martini et al.(2001)]{mar2} Martini, P., Pogge, R. W., Ravindranath, 
   S., \& An, J. H. 2001, \apj, in press
\bibitem[Mirabel et al.(1999)]{mir} Mirabel, I. F., Laurent, O., Sanders, D. B.,
   Sauvage, M., Tagger, M., Charmandaris, V., Vigroux, L., Gallais, P.,
   Cesarsky, C., \& Block, D. L. 1999, \aap, 341, 667
\bibitem[Moles et al.(1995)]{mol} Moles, M., M\'arquez, I., \& P\'erez, E. 1995, 
   \apj, 438, 604
\bibitem[M\"{o}llenhoff et al.(1995)]{moll} M\"{o}llenhoff, C., Matthias, M.,
   \& Gerhard, O. E. 1995, \aap, 301, 359
\bibitem[Mulchaey \& Regan(1997)]{mul} Mulchaey, J. S., \& Regan, M. W. 1997,
   \apjl, 482, L135
\bibitem[Noguchi(1988)]{nog} Noguchi, M. 1988, \aap, 203, 259 
\bibitem[Peletier et al.(1999)]{pel} Peletier, R. F., Balcells, M., Davies, 
   R. L., Andredakis, Y., Vazdekis, A., Burkert, A., \& Prada, F. 1999, \mnras,
   310, 703
\bibitem[Peletier et al.(1999)]{pel99} Peletier, R. F., Knapen, J. H.,
   Shlosman, I., P\'erez--Ram\'\i rez, D., Nadeau, D., Doyon, R., Rodriguez
   Espinosa, J. M., \& P\'erez Garc\'\i a, A. M. 1999, \apjs, 125, 363
\bibitem[Peletier et al.(1995)]{pel2} Peletier, R. F., Valentijn, E. A., 
   Moorwood, A. F. M., Freudling, W., Knapen, J. H., \& Beckman, J. E. 1995,
   \aap, 300, L1
\bibitem[Pfenniger \& Norman(1990)]{pfe} Pfenniger, D., \& Norman, C. 1990,
   \apj, 363, 391
\bibitem[Quillen et al.(1999)]{qui99} Quillen, A. C., Alonso-Herrero, A., 
   Rieke, M. J., Rieke, G. H., Ruiz, M., \& Kulkarni, V. 1999, \apj, 527, 696
\bibitem[Regan \& Mulchaey(1999)]{reg1} Regan, M. W., \& Mulchaey, J. S. 1999, 
   \aj, 117, 2676
\bibitem[Rieke \& Lebofsky(1985)]{riek} Rieke, G. H., \& Lebofsky, M. J. 1985,
   \apj, 288, 618
\bibitem[Salo(1991)]{salo} Salo, H. 1991, \aap, 243, 118
\bibitem[Sandage \& Brucato(1979)]{san} Sandage, A., \& Brucato, R. 1979, 
   \aj, 84, 472
\bibitem[Schwarz(1984)]{scw} Schwarz, M. P. 1984, \mnras, 209, 93 
\bibitem[Schmitt(2001)]{sch} Schmitt, H. R. 2001, \aj, in press 
\bibitem[Scoville et al.(2000)]{sco20} Scoville, N. Z. et al. 2000, \aj, 
   119, 991
\bibitem[Seigar et al.(2000)]{sei} Seigar, M. Carollo, M., Dejonghe, H.,
   Stiavelli, M., \& de Zeeuw, T. 2000, in ASP Conf. Ser. 197, Dynamics of 
   Galaxies: from the Early Universe to the Present, ed. G. A. Mamon, \& 
   V. Charmandaris (San Francisco: ASP), 269 
\bibitem[Shaw et al.(1993)]{sha1} Shaw, M. A., Combes, F., Axon, D. J., \& 
   Wright, G.S. 1993, \aap, 273, 31
\bibitem[Shaw et al.(1995)]{sha2}  Shaw, M. A., Axon, D. J., Probst, R., 
   \& Gatley, I. 1995, \mnras, 274, 369
\bibitem[Sheth et al.(2000)]{she} Sheth, K., Regan, M. W., Vogel, S. N., \& 
   Teuben, P.J. 2000, \apj, 532, 221  
\bibitem[Shlosman et al.(1989)]{shl1} Shlosman, I., Frank, J., \& Begelman, M. C.
   1989, \nat, 338, 45
\bibitem[Shlosman et al.(2000)]{shl20} Shlosman, I., Peletier, R. F., \& Knapen,
   J. H. 2000, \apjl, 535, L83
\bibitem[Shlosman \& Heller(2001)]{shl21} Shlosman, I., \& Heller, C. H. (2001),
   \apj, in press (preprint astro-ph/0109536)
\bibitem[Simkin et al.(1980)]{sim} Simkin, S. M., Su, H. J., \& Schwarz, M. P.
   1980, \apj, 237, 404
\bibitem[Thompson \& Corbin(1999)]{tho} Thompson, R. I., \& Corbin, M. 1999,
   \apss, 266, 79
\bibitem[Tully(1988)]{tul} Tully, R.B. 1988, Nearby Galaxies Catalog, 
   (Cambridge: Cambridge University Press)
\bibitem[van den Bosch \& Emsellem(1998)]{bos} van den Bosch, F. C., \& Emsellem,
   E. 1998, \mnras, 298, 267
\bibitem[Veron et al.(1981)]{ver1} Veron, M. P., Veron, P. \& Zuiderwijk, E. J.
   1981, \aap, 98, 34
\bibitem[Veron-Cetty \& Veron(1991)]{ver2}Veron-Cetty, M.--P., \& Veron, P. 1991,
   A Catalogue of Quasars and Active Nuclei, (5th ed.; Garching: ESO)
\bibitem[Wozniak et al.(1995)]{woz} Wozniak, H., Friedli, D., Martinet, L.,
   Martin, P., \& Bratschi, P. 1995, \aaps, 111, 115
\end{thebibliography}
\end{document}